\documentclass[prd,nofootinbib,showpacs]{revtex4}
\usepackage{graphics}
\usepackage{bm}
\usepackage{graphicx}
\usepackage{amsmath}
\usepackage{amssymb}
\usepackage{rotating,multirow}
\usepackage{color}

\def\be{\begin{equation}}
\def\ee{\end{equation}}
\def\ba{\begin{eqnarray}}
\def\ea{\end{eqnarray}}
\def\l{\left}
\def\r{\right}
\def\hub{{\cal H}}
\def\f{\frac}

\def\etal{{\frenchspacing\it et al.}}
\def\ie{{\frenchspacing\it i.e.}}

\def\etc{{\frenchspacing\it etc.}}

\def\lama{{\rm log}(\lambda_{1}^{2}/{\rm Mpc}^2)}
\def\lamb{{\rm log}(\lambda_{2}^{2}/{\rm Mpc}^2)}

\begin{document}

\title{Searching for modified growth patterns with tomographic surveys}

\author{Gong-Bo Zhao$^{1}$, Levon Pogosian$^{1}$, Alessandra Silvestri$^{2,3}$, and Joel Zylberberg$^{1,4}$}

\smallskip
\affiliation{$^1$Department of Physics, Simon Fraser University, Burnaby, BC, V5A 1S6, Canada \\
\smallskip
$^2$Department of Physics, Syracuse University, Syracuse, NY 13244, USA \\
\smallskip
$^3$ Kavli Institute for Astrophysics and Space Research, MIT,
Cambridge, MA 02139, USA  \\
\smallskip
$^4$ Department of Physics, University of California, Berkeley, CA 94720, USA}

\begin{abstract}
In alternative theories of gravity, designed to produce cosmic acceleration at the current epoch,
the growth of large scale structure can be modified. We study the potential of upcoming and future
tomographic surveys such as DES and LSST, with the aid of CMB and supernovae data,
to detect departures from the growth of cosmic structure expected within General Relativity.
We employ parametric forms to quantify the potential time- and scale-dependent variation of the effective
gravitational constant, and the differences between the two Newtonian potentials.
We then apply the Fisher matrix technique to forecast the errors on the modified growth
parameters from galaxy clustering, weak lensing, CMB, and their cross-correlations across multiple
photometric redshift bins. We find that even with conservative assumptions about the data,
DES will produce non-trivial constraints on modified growth, and that LSST will do significantly better.
\end{abstract}


\date{\today}

\maketitle

\section{Introduction}
\label{Intro} Observations strongly favor a universe that has
recently entered a phase of accelerated
expansion~\cite{Riess:1998cb,perlmutter}. This poses a puzzle for
modern cosmology as standard \emph{General Relativity} (GR), applied
to a universe which contains only radiation and dust, has
difficulties fitting the data. One can view this as evidence for the
existence of \emph{Dark Energy} (DE) -- a yet unknown component with
a negative equation of state, such as a cosmological constant,
$\Lambda$. An alternative explanation could involve modifying GR in
a manner that leads to accelerating solutions.  Popular examples
include the so-called $f(R)$ class of
models~\cite{Starobinsky:1980te,Capozziello:2003tk,Carroll:2003wy,Starobinsky:2007hu,Nojiri:2008nk,Boisseau:2000pr},
Chameleon type scalar-tensor theories~\cite{Khoury:2003aq}, the
\emph{Dvali-Gabadadze-Porrati} (DGP) model~\cite{Dvali:2000hr}, and
models motivated by DGP, such as the recently introduced
Degravitation scenario~\cite{Dvali:2007kt}.

Although the $\Lambda$CDM model, consisting of GR  with $\Lambda$ and \emph{Cold Dark Matter} (CDM), is currently the best fit to the data, it faces some challenges on the theoretical side, such as the coincidence and the fine-tuning problems. On the other side, alternative theories of gravity have to satisfy the multitude of existing experimental tests passed by GR~\cite{Will,Will:2005va}. This typically requires a degree of fine-tuning which, at best, does not improve on that involved in setting $\Lambda$ to the value required by current data. Moreover, once these modified theories are tuned to avoid conflicts with existing constraints (when it is possible), their predictions for the expansion history of the universe are often identical to that of the $\Lambda$CDM model\footnote{The modifications of gravity discussed in this work do not attempt to replace dark matter. We assume existence of CDM.}\cite{Dvali:2007kt,Hu:2007nk,Appleby:2007vb,Pogosian:2007sw,Brax:2008hh,Capozziello:2008fn}. However, this degeneracy is typically broken at the level of cosmological structure formation; indeed, models of modified gravity that closely mimic the cosmological constant at the background level can still give significantly different predictions of the growth of structure.  The large scale structure of the universe therefore offers a promising testing ground for GR and it is important to explore to what extent one can detect departures from GR in the growth of structure with present and upcoming cosmological data.

By definition, the term `modified gravity' implies that the form of the Einstein-Hilbert action is different from that of GR, and, as a consequence, the Einstein equations are changed. At the background level, the modifications allow for a late-time acceleration, which is typically degenerate with $\Lambda$CDM after the required tuning. However, since the equations describing the evolution of cosmological perturbations are modified as well, models with the same expansion history as in $\Lambda$CDM can lead to different dynamics for the growth of cosmic structure.

Structure formation has been studied for f(R) models in~\cite{Song:2006ej, Bean:2006up, Pogosian:2007sw,Tsujikawa:2007xu}, for Chameleon models in~\cite{Brax:2005ew}, and for the DGP model in~\cite{Koyama:2005kd,Song:2006jk,Song:2007wd,Cardoso:2007xc}. One way to test the consistency of the $\Lambda$CDM model is to compare values of the cosmological parameters extracted from distance measures, such as supernovae magnitudes and baryon acoustic oscillations, to the values found from growth measures, such as galaxy counts and weak lensing \cite{Ishak:2005zs}.
One can also introduce general parametrizations of the modified evolution of gravitational potentials and matter perturbations~ \cite{Zhang:2007nk,Amin:2007wi,Hu:2007pj,Hu:2008zd,Bertschinger:2008zb} for the purpose of detecting/constraining departures from GR.

Scalar metric perturbations in the Newtonian gauge are described by two potentials, $\Psi(\vec{x},t)$ and $\Phi(\vec{x},t)$, which correspond to perturbations in the time-time and space-space components of the metric tensor, respectively. In the $\Lambda$CDM model, these two potentials are equal during the epoch of structure formation, and their time dependence is set by the same scale-independent linear growth function $g(a)$ that describes the growth of matter density perturbations. This, generally, is no longer true in theories of modified gravity, where one can have scale-dependent growth patterns. The two Newtonian potentials need not be the same, and their dependence on matter perturbations can be different. Working in Fourier space, we parametrize the ratio between the Newtonian potentials and the dependence of $\Psi$ on the matter density perturbation with two time- and scale-dependent functions $\gamma(a,k)$ and $\mu(a,k)$. We then study the potential of upcoming and future tomographic surveys to constrain departures of these functions from their GR values. Dealing with unknown functions implies working with an infinite number of degrees of freedom. Similarly to the more widely studied problem of determining the dark energy equation of state, $w(z)$, one can proceed in several ways:
\begin{enumerate}
\item One can assume a functional form motivated by a certain class of theories, with a few parameters, and forecast the constraints on the parameters. This would tell us the extent to which one can constrain these theories and also allow us to reconstruct the shape of the functions based on the chosen form. The results, of course, would depend on the choice of the parametrization. However, they would be good indicators of the power of current and upcoming surveys to constrain departures from GR. Moreover, as we will discuss shortly, it is possible to employ a parametrization that accurately represents a broad class of modified theories. This is the approach we take in this paper.
\item Another, non-parametric approach, consists of performing a \emph{Principal Component Analysis} (PCA) to determine the eigenmodes of $\mu$ and $\gamma$ that can be constrained by data~\cite{PCA1,PCA2}. This method
allows one to compare different experiments and their combinations according to the relative gain in information about the functions. PCA can also point to the ``sweet spots'' in redshift and scale where data is most sensitive to variations in $\mu$ and $\gamma$, which can be a useful guide for designing future observing strategies.  The PCA method does not allow one to reconstruct the shape of the functions from data. However, one can still reproduce the errors on parameters of any parametrization from the eigenvectors and eigenvalues found using PCA~\cite{PCA2}. Hence, in terms of forecasting the errors, the PCA method can do everything that the first method can, plus the benefits mentioned above. It is, however, more demanding computationally, and one needs a criterion for deciding which modes are well-constrained. We consider the PCA method in a separate publication \cite{modgrav_pca}.
\item A third approach, which can work for certain estimators of $\gamma$, is a direct reconstruction from data. In~\cite{Zhang:2007nk,Zhang:2008ba}, it was proposed to consider the ratio of the peculiar velocity-galaxy correlation with the weak lensing - galaxy correlation. In such a ratio, the dependence on the galaxy bias cancels out. Then, since the peculiar velocities are determined by the potential $\Psi$, while the weak lensing is controlled by $\Phi+\Psi$, such ratio, if appropriately constructed, would directly probe any difference between $\Phi$ and $\Psi$. This is a more direct and model-independent way of testing GR with the growth of structure than the first two methods. Its power, however, will depend on how well future experiments will be able to measure  peculiar velocities. Also, while it may allow the extraction of $\gamma$, it does not directly probe $\mu$. Still, this is a novel and promising method that should be pursued in parallel with the first two.
\end{enumerate}

Our parametric forms for $\gamma(a,k)$ and $\mu(a,k)$ are analogous
to those introduced in~\cite{Bertschinger:2008zb}, and contain a
total of five parameters. These forms are highly accurate in
describing the linear growth in a wide class of scalar-tensor
theories and also flexible enough to capture features of modified
dynamics in other theories. We use the Fisher matrix technique to
forecast the errors on the modified growth parameters, along with
the standard set of cosmological parameters. We represent the data
in terms of a set of all possible two-point correlation functions
(both auto- and cross-correlations) between the galaxy counts, weak
lensing shear, and \emph{Cosmic Microwave Background} (CMB)
temperature anisotropy, across multiple redshift bins, in addition
to the CMB E-mode polarization autocorrelation, and the CMB E-mode
and temperature cross-correlation. We find that even with a
conservative treatment of data, such as using only the modes that
are well within the linear regime, the upcoming and future
tomographic surveys, like the \emph{Dark Energy Survey} (DES) and
\emph{Large Synoptic Survey Telescope} (LSST), in combination with
CMB and future \emph{SuperNovae} (SNe) luminosity-distance data,
will be able to produce non-trivial constraints on all five
parameters. We also show that LSST will significantly improve our
ability to test GR, compared to DES.

The rest of the paper is organized as follows. In Sec.~\ref{theory}, we motivate our choice of parametric forms for $\gamma(a,k)$ and $\mu(a,k)$. In Sec.~\ref{observables}, we describe the set of observables expected from tomographic weak lensing surveys, their dependence on the underlying gravitational potentials, and the Fisher matrix technique we used to forecast the parameter errors. We also describe the experiments considered -- DES~\cite{DES}, LSST~\cite{LSST}, the \emph{SuperNova/Acceleration Probe} (SNAP)~\cite{SNAP}, and Planck~\cite{Planck}. We present our results in Sec.~\ref{results} along with a discussion of their dependence on the assumptions made in the analysis. We conclude with a summary in Sec.~\ref{summary}.

\section{Parametrizing the modifications of gravity}\label{theory}
We study the evolution of linear matter and metric perturbations in a general metric theory of gravity. Assuming that the background evolution is correctly described by the flat \emph{Friedmann-Robertson-Walker} (FRW) metric, we focus on scalar perturbations and work in the conformal Newtonian gauge, so that the perturbed line element is given by
\be\label{metric}
ds^2=-a^2(\eta)\l[\l(1+2\Psi(\vec{x},\eta)\r)d\eta^2-\l(1-2\Phi(\vec{x},\eta)\r)d\vec{x}^2\r] \ ,
\ee
where $\eta$ is the conformal time. In the rest of the paper, all the perturbed quantities are presented in Fourier space, dots indicate derivatives with respect to $\eta$, and $\hub\equiv \dot{a}/a$ (as opposed to $H$, which contains a derivative with respect to cosmic time $t$, $H\equiv a^{-1}da/dt$). We use the standard notation for the energy momentum tensor of the matter fields, which to first order in the perturbations, assumes the following form
\ba\label{en-mom_tensor}
&&{T^0}_0=-\rho(1+\delta),\nonumber\\
&&{T^0}_i=-(\rho+P)v_i,\\
&&{T^i}_j=(P+\delta P)\delta^i_j+\pi^i_j,
\ea
where $\delta\equiv \delta\rho/\rho$ is the density contrast, $v$ the velocity field, $\delta P$ the pressure perturbation and $\pi^i_j$ denotes the traceless component of the energy-momentum tensor. Finally, we define the anisotropic stress $\sigma$ via $(\rho+P)\sigma\equiv -(\hat{k}^i\hat{k}_j-\f{1}{3}\delta^i_j)\pi^i_j$.

In GR, the linearized Einstein equations provide two independent equations relating the metric potentials and matter perturbations, the Poisson and anisotropy equations, respectively:
\ba\label{Poisson}
&&k^2\Phi=-\f{a^2}{2M_P^2}\rho\Delta \ ,\\
\label{anisotropy}
&&k^2(\Phi-\Psi)=\f{3a^2}{2M_P^2}(\rho+P)\sigma \,,
\ea
where $\rho\Delta\equiv\rho\delta+3\f{aH}{k}(\rho+P)v$ is the comoving density perturbation.
In the $\Lambda$CDM and minimally coupled quintessence models, the anisotropic stress is negligible at times relevant for structure formation, and we have $\Psi=\Phi$.

In models of modified gravity, as well as in more exotic models of dark energy, the relation between the two Newtonian potentials, and between the potentials and matter perturbations, can be different~\cite{Zhang:2005vt,Bertschinger:2006aw,Song:2006ej}. We parametrize the changes to the Poisson and the anisotropy equations as follows:
\ba\label{parametrization-Poisson}
&&k^2\Psi=-\f{a^2}{2M_P^2}\mu(a,k)\rho\Delta\\
\label{parametrization-anisotropy}
&&\f{\Phi}{\Psi}=\gamma(a,k) \ ,
\ea
where $\mu(a,k)$ and $\gamma(a,k)$ are two time- and scale-dependent functions encoding the modifications of gravity and/or the contribution of an exotic dark energy fluid. Note that we have chosen to define $\mu$ via the Poisson equation (\ref{parametrization-Poisson}) written in terms of $\Psi$, the perturbation to the time-time component of the metric. This choice is natural, as it is $\Psi$ that enters the evolution equation for CDM density perturbations on sub-horizon scales\footnote{Our rescaling of the Newton's constant $\mu$ corresponds to $G_{\Phi}/G$ in the notation of~\cite{Bertschinger:2008zb} -- their $\Phi$ is our $\Psi$, and {\it vice versa}.}:
\be
\label{density_evol}
\ddot{\delta}+\hub\dot{\delta}+k^2\Psi=0 \ .
\ee

The gravitational potentials are not going to be observed directly. In order to see the effect of the modifications on observable quantities, such as \emph{Galaxy Counts} (GC), \emph{Weak Lensing} shear (WL) and CMB, we modify the publicly available {\it Code for Anisotropies in the Microwave Background} (CAMB)~\cite{camb,Lewis:1999bs}. This requires implementing Eqs.~(\ref{parametrization-Poisson}) and (\ref{parametrization-anisotropy}) in the synchronous gauge used in CAMB. The details of this procedure are given in Appendix~\ref{camb_implementation}, where we also demonstrate that our method respects the super-horizon consistency condition~\cite{Wands:2000dp,Bertschinger:2006aw}.

In addition to the effects of modified gravity, the evolution of all cosmological perturbations depends on the background expansion. We restrict ourselves to background histories consistent with the flat $\Lambda$CDM model. One reason for this choice is that $\Lambda$CDM is currently the best fit to available data. Another reason comes from the fact that in the recently popular models of modified gravity, e.g. the scalar-tensor models, $f(R)$, nDGP and Degravitation (but not sDGP), the expansion history is effectively the same as in $\Lambda$CDM, and the main differences between models arise from the evolution of cosmic structure. Finally, by restricting ourselves to $\Lambda$CDM-like expansion histories, we can better distinguish the effects of modified growth. We do not, however, fix the values of the cosmological parameters $\Omega_m h^2$, $\Omega_b h^2$ and $h$, nor the spectral index $n_s$ or the optical depth $\tau$. These cosmological parameters will be varied along with the modified growth parameters. In principle, it would be interesting to also allow for a non-zero spatial curvature and for variations in the effective dark energy equation of state. We have included these effects in our forecasts for $\mu(a,k)$ and $\gamma(a,k)$ based on the PCA approach~\cite{modgrav_pca}.

To arrive at a suitable parametrization of the functions $\mu(a,k)$
and $\gamma(a,k)$, we note that models of modified gravity typically
introduce a transition scale which separates regimes where gravity
behaves differently. For example, in $f(R)$ and scalar-tensor
models, the functions $\mu$ and $\gamma$ are equal to unity at early
times and on large scales ($\ie$ on scales that are larger than the
characteristic scale of the model), and they transition to a
modified value on smaller scales and late times. We can mimic this
time- and scale-dependent transition via the following functions of
the variable $k^2a^s$ \ba\label{par_G}
&&\mu(a,k)=\f{1+\beta_1\lambda_1^2\,k^2a^s}{1+\lambda_1^2\,k^2a^s}\\
\label{par_gamma}
&&\gamma(a,k)=\f{1+\beta_2\lambda_2^2\,k^2a^s}{1+\lambda_2^2\,k^2a^s} \ ,
\ea
where the parameters $\lambda^2_i$ have dimensions of length squared, while the $\beta_i$ represent dimensionless couplings. The expressions~(\ref{par_G}) and~(\ref{par_gamma}) coincide with the scale-dependent parametrization introduced in~\cite{Bertschinger:2008zb}. It is easy to show that this parametrization follows from scalar-tensor theories, of which the $f(R)$ models are an example. In these models, one makes a distinction between the so-called Jordan frame, with the metric $g_{\mu\nu}$, (where matter falls along the geodesics and the action for gravity is modified), and the Einstein frame, with the metric $\tilde{g}_{\mu\nu}$, in which the Einstein form of the action is preserved but there is an additional scalar field non-minimally coupled to gravity~\cite{Cotsakis:1988,Magnano:1993bd,Chiba:2003ir}. In what follows, we adopt the usual convention of indicating Einstein frame quantities with a tilde. The two frames are related through a conformal mapping:
\be\label{mapping}
\tilde{g}_{\mu\nu}=e^{\kappa\alpha_i(\phi)}g_{\mu\nu} \,,
\ee
where $\alpha_i(\phi)$ is defined in Eq.~(\ref{Einstein_action_text}), and the index $i$ indicates different matter components.
In the Einstein frame, the action is a Chameleon-type~\cite{Khoury:2003aq} action for the massive field $\phi$ coupled to matter fields via couplings $\alpha_i(\phi)$
\be\label{Einstein_action_text}
S_E=\int d^4x\sqrt{-\tilde{g}}\l[\f{M_P^2}{2}\tilde{R}-\f{1}{2}\tilde{g^{\mu\nu}}(\tilde{\nabla}_{\mu}\phi)\tilde{\nabla}_{\nu}\phi-V(\phi)\r]+S_i\l(\chi_i,e^{-\kappa\alpha_i(\phi)}\tilde{g}_{\mu\nu}\r)\,.
\ee
For $f(R)$ theories, the coupling is universal and linear, $\alpha=\sqrt{2/3}\,\phi$, however, in a general scalar-tensor theory, the coupling(s) can be a non-linear function(s) of the field $\phi$. In what follows, we will focus on the growth of structure and consider primarily cold dark matter, working with a single coupling $\alpha(\phi)$.

The behavior of cosmological perturbations in general models of coupled dark energy has been studied in detail in~\cite{Bean:2001ys,Amendola:2003wa,Bean:2007ny,Bean:2008ac}. The scalar field does not contribute any anisotropic stress, neither do matter fields, so the Newtonian potentials in the Einstein frame are equal: $\tilde{\Phi}=\tilde{\Psi}$. However, the effective potential acting on dark matter particles has an extra contribution due to the interaction of matter with the field $\phi$~\cite{Amendola:2003wa}. Neglecting baryons and radiation, one obtains the following Poisson equation
\be\label{eff_Poisson}
\Psi_{\rm{eff}}=-\f{3}{2}\f{\tilde{a}^2\tilde{\rho}_m}{k^2}\tilde{\delta}_m\l(1+\f{1}{2}{\alpha'}^2Y(k)\r) \ ,
\ee
where $Y(k)=k^2/(k^2+a^2m^2)$ is the Yukawa term, $m$ is the time-dependent effective mass of the scalar field, and primes denote differentiation w.r.t. the field $\phi$.  Following the mapping prescription described in Appendix~\ref{E_J_mapping}, we can now map these equations back to the Jordan frame.
As a result of this mapping, we obtain the following effective Newton constant $\mu$ and ratio between the potentials $\Phi$ and $\Psi$, $\gamma$:
\ba\label{G_scalartensor}
&&\mu(a,k)=\f{1+\l(1+\f{1}{2}{\alpha'}^2\r)\f{k^2}{a^2m^2}}{1+\f{k^2}{a^2m^2}}\\
\label{gamma_scalartensor}
&&\gamma(a,k)=\f{1+\l(1-\f{1}{2}{\alpha'}^2\r)\f{k^2}{a^2m^2}}
{1+\l(1+\f{1}{2}{\alpha'}^2\r)\f{k^2}{a^2m^2}} \ .
\ea
In writing the expression for $\mu$, we neglected an overall pre-factor of $e^{-\kappa\alpha(\phi)}$, which corresponds to a time-dependent rescaling of the Newton constant. In the specific case of $f(R)$ theories it corresponds to $1+f_R$, where $f_R\equiv df/dR$. The consistency of scalar-tensor theories with local\footnote{Even if $f(R)$ models can be made to satisfy solar system constraints, they may still lead to singularities when one considers formation of compact objects, such as neutron stars~\cite{Appleby:2008tv,Frolov:2008uf,Kobayashi:2008tq}. Avoiding these singularities would require additional fine-tuning.} and cosmological tests requires that $e^{-\kappa\alpha(\phi)}\simeq 1$, with departures from unity being outside the reach of current and upcoming cosmological probes. Therefore, it is safe to approximate it with unity.

The expressions~(\ref{G_scalartensor})~-~(\ref{gamma_scalartensor}) are equivalent to the parametrization~(\ref{par_G})~-~(\ref{par_gamma}) once the parameters $\{s,\lambda_i^2,\beta_i\}$ are chosen as follows
\ba\label{st_parameters}
&&a^{(1+s/2)}=\f{m_0}{m}\nonumber\\
&&\lambda_1^2=\frac{1}{m_0^2}\nonumber\\
&&\lambda_2^2=\frac{1}{m_0^2}\l(1+\f{{\alpha'}^2}{2}\r)\nonumber\\
&&\beta_1=1+\f{{\alpha'}^2}{2}\nonumber\\
&&\beta_2=\f{2-{\alpha'}^2}{2+{\alpha'}^2}\,.
\ea
In deriving Eqs.~(\ref{st_parameters}), we have assumed $\alpha'\simeq \rm{const}$. This is exact in $f(R)$ theories, where the coupling $\alpha(\phi)$ is a linear function of the field. In more general scalar-tensor theories, the coupling can be a non-linear function of the scalar field. However, the value of the field $\phi$, typically, does not change significantly on the time-scales associated with the epoch of structure formation. It is, therefore, reasonable to approximate $\alpha'$ with a constant for our purposes.

From Eqs.~(\ref{st_parameters}), it is clear that in the case of
scalar-tensor theories, the parameters $\{\lambda_i^2,\beta_i\}$ are
related  by
\be\label{st_consistency}
\beta_1=\f{\lambda_2^2}{\lambda_1^2}=2-\beta_2\f{\lambda_2^2}{\lambda_1^2}\,
. \ee
By fitting forms (\ref{par_G}) and (\ref{par_gamma}) to data and
checking the validity of the relations (\ref{st_consistency}), one
can test and potentially rule out a large class of scalar-tensor
models of cosmic acceleration.

In our forecasts, we will need to
assume particular fiducial values of $\{s,\lambda_i^2,\beta_i\}$.
One type of fiducial cases we consider are based on $f(R)$ theories. The other type is motivated by the Chameleon scenario. In any scalar-tensor theory, the parameter $s$ is determined by the time-evolution of the mass of the scalar. The non-minimal coupling
leads to the dependence of the effective mass of the
scalar on the local energy density of
non-relativistic matter. The effective potential for the scalar
field is
\be\label{eff_potential_Chameleon}
V_{\rm{eff}}(\phi)=V(\phi)+\bar{\rho}_me^{\kappa\alpha(\phi)} \ ,
\ee
which gives an effective mass \be\label{mass_scalar}
m^2=V''_{\rm{eff}}(\phi_{\rm{min}})=V''-\kappa\l(\alpha''+{\alpha'}^2\r)V'
\ , \ee calculated at the minimum $\phi_{\rm{\min}}$ of the
potential $V_{\rm{eff}}$. Following~\cite{Brax:2004qh}, the time
dependence of the mass~(\ref{mass_scalar}) can be approximated as
\be\label{mass_st}
\f{\dot{m}}{m}\approx\f{1}{2}\f{V'''}{V''}\,\dot{\phi}_{\rm{min}}\,.
\ee
For tracking-type potentials described by an inverse
power-law, $\ie$ $V\sim \phi^{-n}$, the mass
evolves as
\be\label{mass_st2} m\sim a^{-3(n+2)/2(n+1)}
\ee
corresponding\footnote{It is easy to show that $s$ must be in this range for any $V(\phi)\sim \phi^a$, for both positive and negative $a$.} to $1<s<4$. Cosmologically viable $f(R)$ models correspond to $s\simeq 4$. The scalar degree of freedom there is represented by the function $f_R$ and its mass is a time dependent function set by $m^2\approx
f_{RR}^{-1}$~\cite{Hu:2007nk,Pogosian:2007sw}. From numerical
simulations~\cite{Pogosian:2007sw}, we find that $f_{RR}\propto
a^{6}$, corresponding to $s\simeq 4$. This can be understood
analytically using the following argument. A viable $f(R)$ must be a
fairly slowly-varying function of R, and in the relevant range of R
we can always approximate it with a power law: $f(R) \sim a_1 + a_2
R^n$ up to the leading order in $R$. This function gives $f_{RR}
\simeq n(n-1) f/R^2 \simeq \rm{const}\cdot R^{-2}$. The mass of the
field is calculated at the minimum of the scalaron potential, where
$R=\kappa^2\rho$, therefore, we have $f_{RR} \propto \rho^{-2}
\propto a^6$, which is what we have found numerically. In the case of tracking-type quintessence potentials, usually considered in Chameleon scenarios, one needs small integer values of the index $n\geq O(1)$ in order to have the appropriate mass scale~\cite{Brax:2004qh}. Therefore, $s\sim 2$ can be seen as a typical value for Chameleon models.

The fiducial values of $\beta_1$ and $\beta_2$ can be derived from $\alpha'$ using Eqs.~(\ref{st_parameters}). $f(R)$ models correspond to a linear coupling with $\alpha'=\sqrt{2/3}$. In this case, Eqs.~(\ref{st_parameters}) give
\ba\label{f_R_parameters1}
&&\beta_1=\f{4}{3}, \ \beta_2=\f{1}{2} \ .
\ea
In Chameleon models, $\alpha'$ is a free parameter
that is typically assumed to be $\sim O(1)$.

Finally, the mass scale today, $m_0$, is a free parameter in all scalar-tensor models.
In the $f(R)$ case, it is constrained
from below by requirements of consistency with local and
cosmological tests. In particular, local tests of
gravity~\cite{Hu:2007nk} set a lower bound, $m_0\gtrsim 10^{-1}$
$\rm{Mpc}^{-1}$, which corresponds to
$\lambda_1^2\lesssim 10^2$ $\rm{Mpc}^2$. This bound can be relaxed
to $m_0\gtrsim10^{-3}$ $\rm{Mpc}^{-1}$, corresponding to
$\lambda_1^2\lesssim 10^6$ $\rm{Mpc}^2$, if we consider only
cosmological tests, neglecting local constraints~\cite{Song:2007da}.
In the case of Chameleons, the constraints are somewhat weaker due to the
additional freedom in choosing $\alpha'$. However, if $\alpha'\sim O(1)$,
then the bounds on $m_0$ should be comparable to those in the $f(R)$ case.

In our forecasts, we use two $f(R)$ fiducial cases, corresponding to two choices of the mass scale $m_0$. Namely, in both $f(R)$ cases, we take
$s=4$, $\beta_1=4/3$ and $\beta_2=1/2$, and try two different sets of values
for ($\lambda_{1}^{2},\lambda_{2}^{2}$):  Model~I with $\lambda_{2}^{2}=\beta_{1}\lambda_{1}^{2}=10^3$ $\rm{Mpc}^2$ and Model~II with $\lambda_{2}^{2}=\beta_{1}\lambda_{1}^{2}=10^4$ $\rm{Mpc}^2$. We also consider two fiducial models corresponding to the Chameleon case, both with $s=2$, $\beta_1=9/8$ and $\beta_2=7/9$, which correspond to $\alpha'=0.5$. We then use the same fiducial values for ($\lambda_{1}^{2},\lambda_{2}^{2}$) as in the $f(R)$ case. We refer to these two models as Model~III and Model~IV. In practice, rather that varying dimensionful parameters $\lambda_i^2$, we work with log$(\lambda_i^2/{\rm Mpc}^2)$. To summarize, we consider the following fiducial cases:
\begin{itemize}
\item Model~I, with $s=4$, $\beta_1=4/3$, $\beta_2=1/2$, log$(\lambda_2^2/{\rm Mpc}^2)={\rm log}(\beta_1\lambda_1^2/{\rm Mpc}^2)=3$;
\item Model~II, with $s=4$, $\beta_1=4/3$, $\beta_2=1/2$, log$(\lambda_2^2/{\rm Mpc}^2)={\rm log}(\beta_1\lambda_1^2/{\rm Mpc}^2)=4$;
\item Model~III, with $s=2$, $\beta_1=9/8$, $\beta_2=7/9$, log$(\lambda_2^2/{\rm Mpc}^2)={\rm log}(\beta_1\lambda_1^2/{\rm Mpc}^2)=3$;
\item Model~IV, with $s=2$, $\beta_1=9/8$, $\beta_2=7/9$, log$(\lambda_2^2/{\rm Mpc}^2)={\rm log}(\beta_1\lambda_1^2/{\rm Mpc}^2)=4$.
\end{itemize}

\begin{figure}[t]
\includegraphics[scale=0.5]{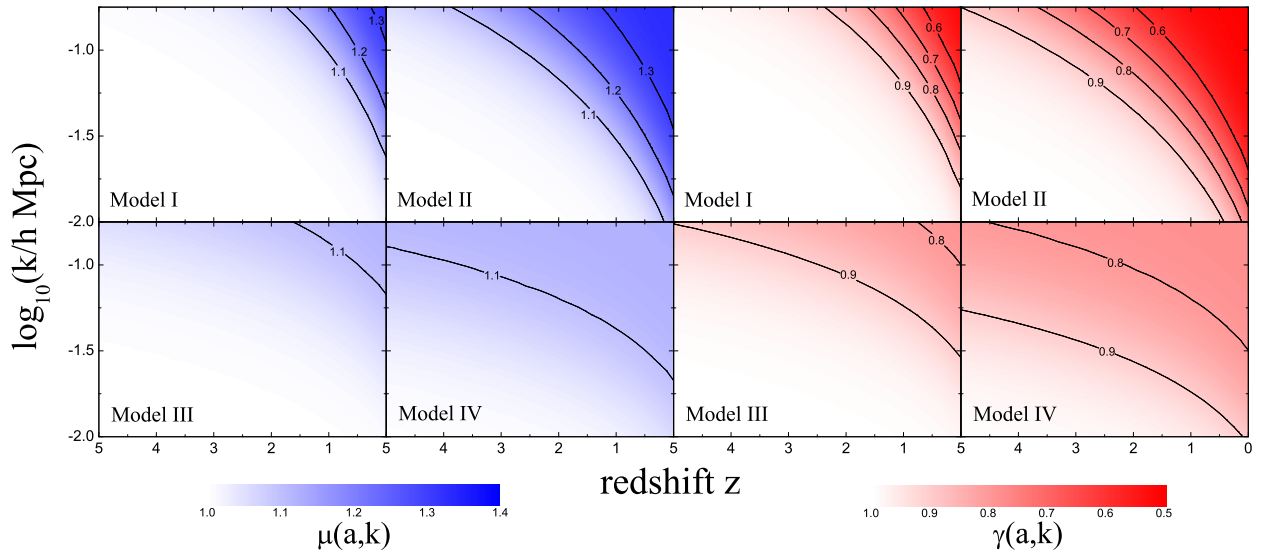}
\caption{The rescaling of the Newton constant $\mu(a,k)$ and the ratio of Newtonian potentials $\gamma(a,k)$, plotted as a function of the redshift for
the four fiducial models used in our Fisher analysis. The contours are lines of constant $\mu$ and $\gamma$. The first two models correspond to $f(R)$ fiducial cases, with $s=4$, the coupling $\beta_1=4/3$ and the mass scale of $\lambda_2^2=10^3 \rm{Mpc}^2$ for model I and $\lambda_2^2=10^4 \rm{Mpc}^2$ for model II. Models III and IV correspond to Chameleon theories with $s=2$, the coupling $\beta_1=9/8$ and the mass scale of $\lambda_2^2=10^3 \rm{Mpc}^2$ and $\lambda_2^2=10^4\rm{Mpc}^2$, respectively. } \label{fig:mu_gamma}
\end{figure}

In Fig.~\ref{fig:mu_gamma}, we plot the functions $\mu(a,k)$ and
$\gamma(a,k)$ corresponding to the four fiducial models considered
in this work. In the $\Lambda$CDM case, both functions are constant
and equal to unity. For Models I-IV, the two functions approach
unity at small values of $k^2a^s$, while at larger $k^2a^s$,
they transition to values determined by $\beta_1$ and $\beta_2$.

In the recent literature, there has been a growing interest in parametrizing modifications of gravity. The parametrization we have used is equivalent to the scale-dependent form proposed in~\cite{Bertschinger:2008zb}. In the same paper, the authors also consider a scale-independent parametrization, in which  $\mu$ and $\gamma$ are not independent functions. A scale-independent parametrization of the ratio of the potentials was also introduced in~\cite{Caldwell:2007cw}, although with a fixed time-dependence. There is, however, no physical model of modified gravity which corresponds to a scale-independent modification of growth. Indeed, to avoid serious conflicts with \emph{Big Bang Nucleosynthesis} (BBN) and CMB data, the modified gravity theories must reduce to GR at early times. They should, however, introduce novel physics at late times on cosmological scales, while obeying local constraints of gravity on non-linear scales. These requirements introduce characteristic transition scales into the theory that lead to a peculiar scale-dependent behavior of perturbations. A parametrization describing all these three regimes, including a possible modified time-dependence on super-horizon scales, is the \emph{Parametrized Post-Friedmann Framework} (PPF) of~\cite{Hu:2007pj}. It consists of three functions of time and space, plus one parameter, plus a parametrization of the non-linear regime. In this work, we are interested in modifications of gravity that primarily become important on sub-horizon cosmological scales. For this purpose, our parametrization, which has five parameters, is sufficient and more economical. That is, our parametrization, by design, reduces to GR on super-horizon scales and evolves into a modified version of gravity only below a certain sub-horizon scale, such as the Compton wavelength of the scalar field. Furthermore, we do not consider the non-linear scales, because describing them properly requires input from $N$-body simulations. This can be done in specific models, such as particular types of $f(R)$~\cite{Oyaizu:2008tb}, but would be ill-defined if one tried to model them based on a general parametrization of linear gravitational potentials, such as Eqs.~(\ref{par_G})
and (\ref{par_gamma}).

The parametrization we use gives an accurate description of the effects of scalar-tensor theories (and in general models of coupled quintessence) on linear scales. Describing models such as DGP would require the introduction of time-dependent modifications on super-horizon scales (and therefore the PPF framework would be a better parametrization in that case). However, we expect that Eqs.~(\ref{par_G})
and (\ref{par_gamma}) can capture some of the features of DGP if applied to transitions on larger scales. In~\cite{Amin:2007wi}, the authors have proposed a framework to test GR on cosmological scales; the parametrization they use is suitable for braneworld theories, while less suited for describing models of the $f(R)$ type. As already mentioned in the introduction, a very promising model-independent procedure to test the relation between the metric potentials which uses the cross-correlation between peculiar velocities and galaxy counts, and galaxy counts with lensing, was introduced in~\cite{Zhang:2007nk}. That method probes $\gamma$ but not $\mu$, and extracting peculiar velocities is notoriously difficult. Finally, in a recent paper~\cite{Bean:2008ac}, the authors considered constraints from CMB and LSS on general theories of interacting dark matter and dark energy.

\section{Tomographic observables}
\label{observables}

In the $\Lambda$CDM model, the sub-horizon evolution of gravitational potentials and the matter density fluctuations are described by a single function of time-- the scale-independent growth factor $g(a)$. Namely, for all Fourier modes that have entered the horizon by some epoch $a_i$, we have $\Delta(a,k)/\Delta(a_i,k)\equiv g(a)$ and the potentials evolve as $\Psi(a,k)/\Psi(a_i,k) =\Phi(a,k)/\Phi(a_i,k)\equiv g(a)/a$. In models of modified gravity, on the other hand, the dynamics of perturbations can be richer and, generically, the evolution of $\Phi$, $\Psi$ and $\Delta$ can be described by different functions of scale and time. By combining different types of measurements, one can try to reconstruct these functions, or at least put a limit on how different they can be. In what follows, we give a brief introduction into the relation between the different types of observables and the gravitational potentials they probe. For a more thorough review of the various ways of looking for modifications in the growth of perturbations we refer the reader to~\cite{Jain:2007yk}.

\emph{Galaxy Counts} (GC) probe the distribution and growth of matter inhomogeneities. However, to extract the matter power spectrum, one needs to account for the bias, which typically depends on the type of galaxies and can be both time- and scale-dependent. On large scales, where non-linear effects are unimportant, one can use a scale-independent bias factor to relate galaxy counts to the total matter distribution. This relation becomes increasingly complicated and scale-dependent as one considers smaller and smaller scales. In principle, the bias parameters can be determined from higher order correlation functions~\cite{fry94,Matarrese:1997sk,Sefusatti:2007ih}. In this work we stick to linear scales where the bias is scale-independent. On sub-horizon linear scales, the evolution of the matter density contrast is determined by Eq.~(\ref{density_evol}). Hence, measurements of GC over multiple redshifts can provide an estimate of $\Psi$ as a function of space and time, up to a bias factor. A more direct probe of the potential $\Psi$, would be a measurement of \emph{peculiar velocities}, which follow the gradients of $\Psi$. Such measurements would be independent of uncertainties associated with modeling the bias. Peculiar velocity surveys typically use redshift-independent distance indicators to separate the Hubble flow from the local flow, and nearby SNeIa are therefore good candidates; a number of surveys, like the 6dFGS~\cite{Jones:2004zy} and the 2MRS~\cite{Erdogdu:2006nd}, use galaxies. An interesting alternative is offered by the kinetic Sunyaev-Zel'dovich effect of clusters~\cite{Sunyaev:1980nv}, that arise from the inverse Compton scattering of CMB photons off high-energy electrons in the clusters. This effect provides a useful way of measuring the bulk motion of electrons in clusters, hence the peculiar velocity of clusters, but it is limited by low signal-to-noise ratio. Current measurements of peculiar velocities are limited in accuracy, and at this point it is not clear how to forecast the accuracy of future observations. Therefore we did not include them in our observables, even though they are a potentially powerful probe~\cite{Song:2008qt}.

In contrast to galaxy counts and peculiar velocities, which respond to one of the metric potentials,  namely $\Psi$, \emph{Weak Lensing} (WL) of distant light sources by intervening structure is determined by spatial gradients of ($\Phi+\Psi$). Hence, measurements of the weak lensing shear distribution over multiple redshift bins can provide an estimate of the space and time variation of the sum of the two potentials. In the $\Lambda$CDM and minimally coupled models of dark energy, the two metric potentials coincide and therefore WL probes essentially the same growth function that controls the evolution of galaxy clustering and peculiar velocities. In models of modified gravity, however, there could be a difference between the potentials, corresponding to an effective shear component, also called a ``gravitational slip''.

Measurements of the \emph{Integrated Sachs-Wolfe} effect (ISW) in the CMB probe the time dependence of the sum of the potentials: $\dot{\Phi}+\dot{\Psi}$. The ISW effect contributes a gain (loss) of energy to CMB photons traveling through decaying (growing) potential wells. It contributes to the CMB anisotropy on the largest scales and is usually extracted by cross-correlation of the CMB with galaxies~\cite{Crittenden:1995ak,Boughn:2003yz,Giannantonio:2008zi}.

By combining multiple redshift information on GC, WL and CMB, and their cross-correlations, one can constrain the differences between the metric potentials and the space-time variation of the effective Newton constant defined in the previous section. Ideally, the experiments would provide all possible cross-correlations, between all possible pairs of observables, in order to maximize the amount of information available to us. In practice, however, it can be difficult to obtain these cross-correlations, since their measurements require that each of the individual fields (CMB, GC, WL) be measured on the same patch of sky. This will be addressed with near and distant future tomographic large scale structure surveys (like DES~\cite{DES}, PAN-STARR~\cite{PAN-STARR} and LSST~\cite{LSST}). We have made separate forecasts using various combinations of data sets, and all possible cross-correlations are considered (CMB-WL, CMB-LSS, LSS-WL). In this manner, we can determine how well modified gravity will be constrained with varying degrees of experimental difficulty and sophistication.

\subsection{Angular spectra}

Consider a two-point correlation function $C^{XY}(\theta) \equiv C^{XY}(|\mathbf{\hat{n}}_1 - \mathbf{\hat{n}}_2|) \equiv \left<X(\mathbf{\hat{n}}_1)Y(\mathbf{\hat{n}}_2)\right> $ between two 2D-fields, $X$ and $Y$, measured by an observer looking at the sky. Here $\mathbf{\hat{n}}$ is a direction on the sky and $\cos{\theta} \equiv \mathbf{\hat{n}}_1 \cdot \mathbf{\hat{n}}_2$. $C^{XY}(\theta)$ can be expanded in a Legendre series:
\begin{equation}
C^{XY}(\theta) = \sum_{\ell=0}^{\infty} \frac{2\ell+1}{4 \pi}C_\ell^{XY} P_\ell(\cos\theta) \ ,
\label{eq:leg_series}
\end{equation}
where the first two terms in the series (the monopole and dipole contributions) are coordinate dependent and should vanish in the CMB frame of a homogeneous and isotropic universe. The expansion coefficients $C_\ell^{XY}$ can be expressed in terms of the primordial curvature power spectrum $\Delta_{\cal R}^{2}$ and the angular transfer functions $I_{\ell}^{X,Y}(k)$ as
\begin{equation}
C_\ell^{XY}= 4\pi \int \frac{dk}{k} \Delta_{\cal R}^{2}
I_{\ell}^X(k) I_{\ell}^Y(k), \label{eq:gen}
\end{equation}
where
\begin{equation}
I_{\ell}^X(k) =  c_{{\cal X}{\cal R}}\int_0^{z_*} d z W_X(z)  j_\ell[kr(z)]\tilde{\cal X}(k, z).
\label{eq:I_gen}
\end{equation}
and similarly for $I_\ell^Y$. A detailed derivation of the above expressions is given in Appendix~\ref{append:cl}. The observable quantities for which we need to evaluate the $I_\ell$'s are the galaxy distributions at several ranges in redshift (the tomographic redshift bins), the maps of lensing shear for different bins, and the CMB temperature anisotropy.

\label{sec:cl}
\begin{figure}[t]
\begin{center}
\includegraphics[scale=0.6]{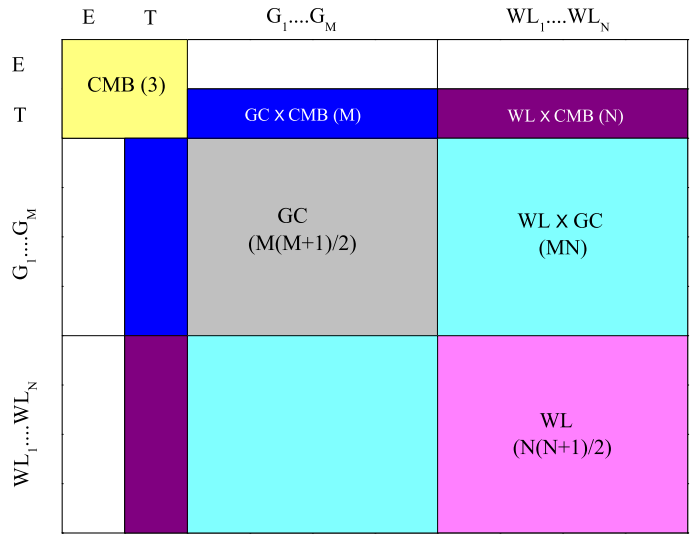}
\caption{A schematic representation of all the $2$-point correlations used in this work. $M$ is the number of redshift bins used for galaxy counts (GC), $N$ is the number of bins used for weak lensing (WL), $T$ and $E$ stand for temperature and E-mode polarization of the CMB. The white blocks corresponding to the cross-correlations between E and G$_i$, and between E and WL$_i$ indicate that those cross-correlations were not considered in this work. }
\label{Fig:table}
\end{center}
\end{figure}

\begin{figure}[t]
\begin{center}
\includegraphics[scale=0.6]{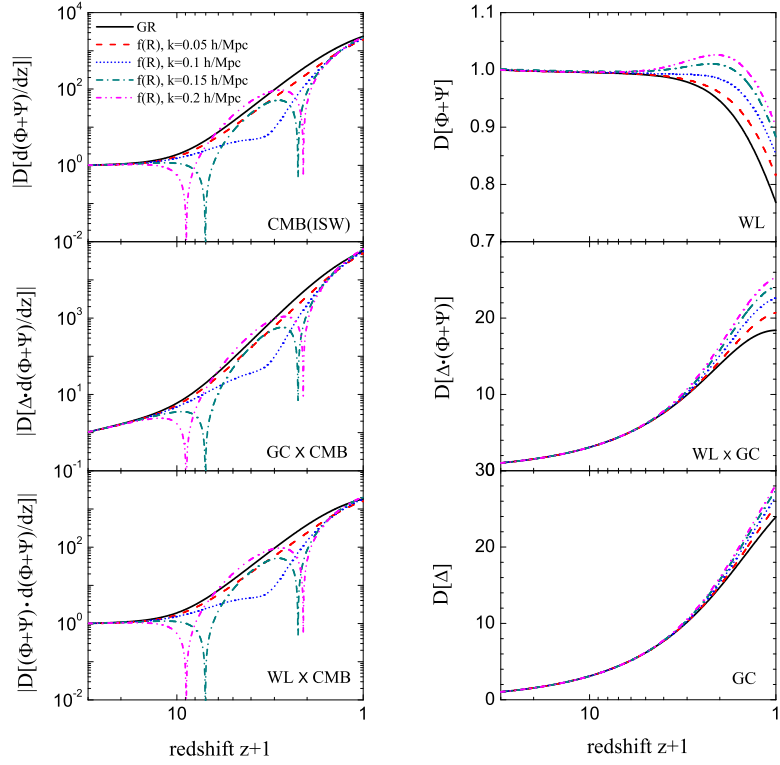}
\caption{Some representative growth functions (time evolution
divided by the corresponding initial values, set at $z=30$) that
contribute to the integration kernels of the $C_{\ell}$'s for ISW,
WL, GC, and their cross-correlations. The corresponding $C_{\ell}$'s
are shown in Fig.~\ref{fig:fid_cl}. The solid line corresponds to GR
and the other lines are for Model II (an $f(R)$ fiducial model
described in Sec.~\ref{theory}). The different types of line
types/colors represent four different $k$ modes and are explained in
the legend.}\label{fig:kernel}
\end{center}
\end{figure}

The distribution of galaxies is expected to trace the distribution of dark matter
up to a bias factor which quantifies selection effects specific to the type and color of the galaxies. The bias factor becomes increasingly scale-dependent on smaller scales and its modeling, e.g. using the halo model~\cite{Kravtsov:2003sg,HuJain}, involves assigning on the order of five parameters to each photometric bin. The halo model assumes the validity of General Relativity and would not be directly applicable for testing modified gravity. Also, the modifications of growth that we are trying to detect would be highly degenerate with the variations of the bias parameters. Higher order correlation functions can, in principle, significantly help in reducing this degeneracy~\cite{fry94,Matarrese:1997sk,Sefusatti:2007ih}. Still, this would require calculating three-point statistics of cosmological perturbations in modified growth models, which is beyond the scope of this paper. Instead, we work under the assumption that on large scales, the bias can be treated as scale-independent and can be modeled with one free parameter $b_i$ for each redshift bin $i$.  With this assumption, the corresponding angular transfer functions can be expressed in terms of the dark matter density contrast as
\begin{equation}
I_{\ell}^{G_i}(k) =  b_i c_{\delta{\cal R}}\int_0^{z_*} d z W_{G_i}(z)  j_\ell[kr(z)]\tilde{\delta}(k, z) \ ,
\label{eq:I_delta}
\end{equation}
where $W_{G_i}(z)$ is the normalized selection function for the $i$th redshift bin, $\tilde{\delta}(k, z)$ is the density contrast transfer function, and $c_{\delta{\cal R}}=-9/10$ for the adiabatic initial conditions~\cite{MB,Garriga} that are assumed throughout this work.

For weak lensing, the relevant $I_\ell$'s are given by
\be\label{wls} I_{l}^{\kappa_{i}}(k)=c_{\Psi{\cal R}}\int_0^{z_*} dz
W_{\kappa_{i}}(z) j_{l}[kr(z)] (\tilde{\Psi}+c_{\Phi
\Psi}\tilde{\Phi}) \ , \ee where $c_{\Psi{\cal R}}=3/5$, $c_{\Phi
\Psi}= (1+2R_\nu/5)$, where $R_\nu \equiv (7N_\nu/8)(4/11)^{4/3}$~\cite{MB}, $N_\nu$ is the number of flavors
of relativistic neutrinos, and $W_{\kappa_{i}}(z)$ is the window
function for the $i$th  bin of sheared galaxies with a normalized
redshift distribution $W_{S_i}(z)$: \be
W_{\kappa_{i}}(z)=\int_z^{\infty} dz' \frac{r(z')-r(z)}{r(z)}
W_{S_i}(z') \ . \ee

\begin{figure}
\begin{center}
\includegraphics[scale=0.45]{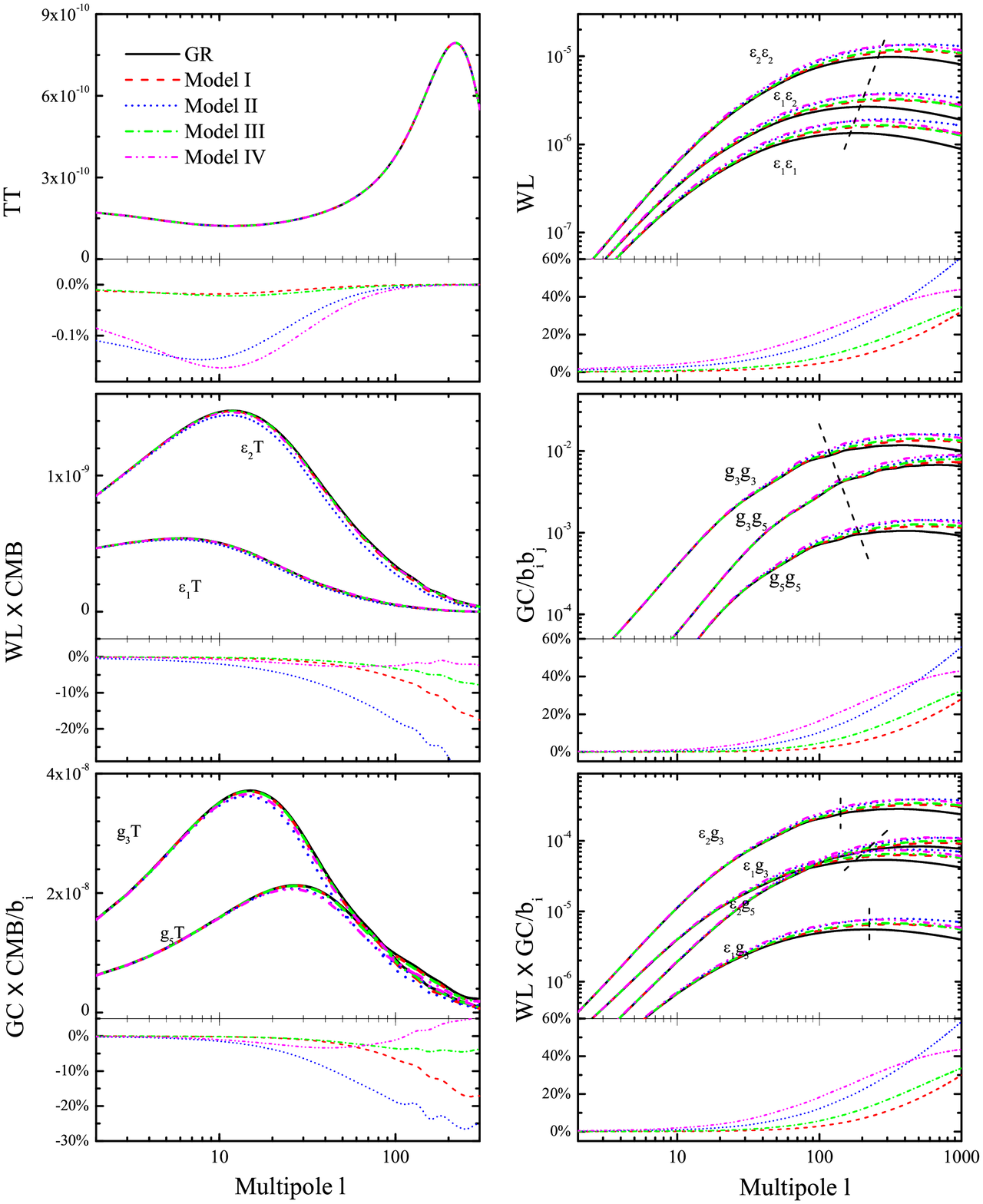}
\caption{Dimensionless power spectra
$\ell(\ell+1)C_{\ell}^{XY}/2\pi$, where $X,Y=\{T,GC,WL\}$ for GR
(solid line) and for our Models I-IV (line types explained in the
legend), for Planck and a few representative choices of GC and WL
redshift bin pairs expected from LSST. On the plots, $\epsilon_i$ stands for the $i$-th WL bin, and
$g_i$ stands for the $i$-th GC bin. The assumed redshift
distribution of WL and GC sources for LSST, along with their
partition into photometric bins, is shown in
Fig.~\ref{Fig:des_lsst_window}. For each group of spectra, we pick one
with the largest deviation from the GR prediction and plot its relative difference w.r.t GR in the lower part of each panel.  The straight dashed lines indicate
the approximate scale at which non-linear corrections become
significant. We have only used the parts of the spectra that can be
accurately described by linear theory.} \label{fig:fid_cl}
\end{center}
\end{figure}

The transfer functions for the CMB temperature anisotropy receive contributions from the last-scattering surface (at $z \sim 1100$) and from more recent redshifts via the Integrated Sachs-Wolfe (ISW) effect. By design, our modifications of gravity should be negligible at recombination. Therefore, their only imprint on the CMB will be via the ISW effect. For the ISW contribution to the CMB, we have
\begin{equation}
I_{\ell}^{ISW}(k) =  c_{\Psi{\cal R}} \int_0^{z_*} dz e^{-\tau(z)} j_{\ell}[kr(z)] {\partial \over \partial z}\left[\tilde{\Psi}+c_{\Phi \Psi}\tilde{\Phi}\right] \ ,
\label{eq:I_ISW}
\end{equation}
where $\tau(z)$ is the opaqueness function.

We numerically evaluate the transfer functions to obtain
$C_{l}^{XY}$ using MGCAMB: a code developed by
us\footnote{Publicly available at
http://www.sfu.ca/$\sim$gza5/MGCAMB.html} for studying modified
growth based on CAMB~\cite{camb}. The details of the implementation
of the parametrization (\ref{parametrization-Poisson}) in CAMB,
which involved conversion to the synchronous gauge, are given in
Appendix~\ref{camb_implementation}. To save computing time, we used
the fact that modifications of GR are negligible at $z>30$, and
until then the usual CAMB solver was used. From $z=30$ and on we
continue by solving Eqs.~(\ref{eq:deltadot}$-$\ref{eq:alphadot}).

As illustrated in Fig.~\ref{Fig:table}, a joint analysis of CMB and
data from a tomographic lensing survey with $M$ GC redshift bins and
$N$ WL bins can give us a total of $3+M(M+1)/2+N(N+1)/2+M+N+MN$
different types of $C_{\ell}$'s from CMB, GC, WL, GC$\times$CMB,
WL$\times$CMB and WL$\times$GC, respectively (we do not correlate
CMB polarization with GC and WL). For example, combining Planck with
DES, with $M=4$ GC bins and $N=4$ WL bins, gives us $47$ different
types of spectra. A combination of Planck with LSST, with $10$ GC
bins and $6$ WL bins, gives us $155$ different $C_{\ell}$'s.

For a given function $f=f(k,z)$ where $k$ and $z$
are comoving wavenumber and redshift respectively, we can define a corresponding growth
function $D$ as
\be\label{eq:growthf}
D[f(k,z)]\equiv\frac{f(k,z)}{f(k,z=30)}.\ee
In Fig.~\ref{fig:kernel}, we plot these growth functions corresponding to the
kernels of several representative $C_{\ell}$'s for ISW, WL, GC and their
cross-correlations. In GR, all of these quantities
grow in a scale-independent manner, while in the modified scenarios considered here
their growth is enhanced in a scale-dependent
way. We see that the enhancement is more pronounced on smaller scales (larger $k$), as expected in scalar-tensor scenarios~\cite{Pogosian:2007sw}.

In Fig.~\ref{fig:fid_cl}, we show some of the spectra expected from
a combination of Planck and LSST
 for the four fiducial models considered in this paper. Shown are the auto- and cross-correlated spectra for two representative WL and GC redshift bins, with central values $z(\varepsilon_{1})\approx 0.3,
z(\varepsilon_{2})\approx 0.8, z(g_{3})\approx 0.4, z(g_{5})\approx 0.9$, where $\epsilon_i$ stands for the $i$-th WL bin, and $g_i$
stands for the $i$-th GC bin.
The full redshift distribution of WL and GC sources we assumed for LSST,
along with their partition into photometric bins, is shown in Fig.~\ref{Fig:des_lsst_window}.

We show only the parts of the spectra that correspond to the linear
cosmological regime. Including higher $\ell$, or smaller scales,
would require us to account for non-linear effects which, strictly
speaking, is not allowed within our framework. To accurately model
growth on non-linear scales, one needs input from N-body
simulations, which can only be performed for specific modified
gravity theories. The fact that we are not testing a specific model,
but constraining a general departure from GR, defined in terms of
linear perturbation variables, precludes us from having a reliable
description of non-linear corrections. Simply applying the
analytical corrections developed under the assumption of GR, such as
by Peacock and Dodds~\cite{Peacock:1996ci} and Smith et
al~\cite{Smith:2002dz}, can lead to significant
errors~\cite{Oyaizu:2008tb}. For this reason, we restrict ourselves
to the linear regime by cutting off the $C_{l}^{XY}$ spectra at
$l_{\rm max} \sim 0.2~h\,\chi(z_{s})$, where $\chi(z_{s})$ is the
comoving distance to the redshift bin at $z=z_{s}$. This cutoff
roughly corresponds to $k\sim 0.2\,h {\rm Mpc}^{-1}$ at
$z=0$. There is certainly a wealth of information about MG parameters on smaller scales. For comparison, using
$k\sim 0.1\,h {\rm Mpc}^{-1}$ at $z =0$ as a cutoff degrades the LSST constraints on MG parameters by a factor of $\sim 2$. However, while it would be tempting to include information from even smaller scales, as mentioned above, it would make predictions obtained using linear theory unreliable.

\subsection{Fisher matrices}
\label{sec:fisher}

In order to determine how well the surveys will be able to constrain
our model parameters, we employ the standard Fisher matrix
technique~\cite{Fisher}. The inverse of the Fisher matrix $F_{ab}$
provides a lower bound on the covariance matrix of the model
parameters via the Cram$\acute{\rm e}$r-Rao inequality, ${\bf C}
\geq {\bf F}^{-1}$~\cite{Fisher}. For zero-mean Gaussian-distributed
observables, like the angular correlations $C^{XY}_\ell$ introduced
in Section~\ref{sec:cl}, the Fisher matrix is given by \be F_{ab} =
f_{\rm sky} \sum_{\ell=\ell_{\rm min}}^{\ell_{\rm max}}\frac{2\ell +
1}{2} {\rm Tr}\left( \frac{\partial {\bf C_\ell}}{\partial p_a} {\bf
\tilde{C}_\ell^{-1}}\frac{\partial {\bf C_\ell}}{\partial p_b} {\bf
\tilde{C}_\ell^{-1}} \right) \ , \label{eq:Fisher} \ee where $p_{a}$
is the ${a}^{\rm th}$ parameter of our model and ${\bf
\tilde{C}_\ell}$ is the ``observed'' covariance matrix with elements
$\tilde{C}^{XY}_\ell$ that include contributions from noise: \be
\tilde{C}^{XY}_\ell= C^{XY}_\ell+N^{XY}_\ell \ . \label{eq:NoiseAdd}
\ee The expression~(\ref{eq:Fisher}) assumes that all fields
$X(\hat{\bf n})$ are measured over contiguous regions covering a
fraction $f_{\rm sky}$ of the sky. The value of the lowest multipole
can be approximately inferred from $\ell_{\rm min} \approx \pi
/(2f_{\rm sky})$. It is also possible to write expressions for
separate contributions to the Fisher matrix from particular subsets
of observables. For example, for angular spectra $C^{X_iX_j}_\ell$,
corresponding to $N$ fixed pairs of fields $(X_i,X_j)$ we can write
\be
\label{eq:subFisher}
F^{\rm sub}_{ab} = f_{\rm sky}
\sum_{\ell=\ell_{\rm min}}^{\ell_{\rm max}} (2\ell + 1)
\sum_{\{ij\}=1}^{N}\sum_{\{mn\}=1}^{N} \frac{\partial
\tilde{C}^{X_iX_j}_\ell}{\partial p_a} [{\bf \tilde{C}^{\rm
sub}_\ell}]^{-1}\frac{\partial \tilde{C}^{X_mX_n}_\ell} {\partial
p_b}  \ ,
\ee
where the covariance matrix ${\bf \tilde{C}^{\rm
sub}_\ell}$ has elements~\cite{HuJain}
\be
[{\bf \tilde{C}^{\rm sub}_\ell}]^{\{ij\}\{mn\}} =
\tilde{C}^{X_iX_m}_\ell \tilde{C}^{X_jX_n}_\ell +
\tilde{C}^{X_iX_n}_\ell \tilde{C}^{X_jX_m}_\ell \ .
\ee
Eqs.~(\ref{eq:Fisher}) and (\ref{eq:subFisher}) become the same in
the limit of summing over all possible pairs $(X_i,X_j)$.

In principle, the noise matrix $N^{XY}_\ell$ includes the expected
systematic errors. Systematics are, however, notoriously difficult
to predict, and are often ignored in parameter constraint forecasts.
For both surveys considered in this work, DES and LSST, achieving
their science goals requires bringing the systematic errors under
the noise level~\cite{DES,Ivezic:2008fe}. Therefore, we take
$N^{XY}_\ell$ to represent statistical noise only. Assuming
uncorrelated Poisson noise on the galaxy overdensity in each bin
($G_i$) and shear fields ($\kappa_i$), the noise is given
by~\cite{HuJain}
\begin{eqnarray}
&&N^{\kappa_i \kappa_j}_\ell = \delta_{ij} \frac{\gamma^2_{rms}}{n_j}   \nonumber \\
&&N^{G_i G_j}_\ell = \delta_{ij} \frac{1}{n_j}  \nonumber \\
&&N^{G_i \kappa_j}_\ell = 0,
\label{eq:Noise}
\end{eqnarray}
where $\gamma_{rms}$ is the expected root mean square shear of the galaxies, and $n_j$ is the number of galaxies per steradian in the $j$th redshift bin.

For CMB, the temperature anisotropy T and the E-mode polarization
are measured in several frequency channels. When combined, the total
noise is less than that in any of the channels
\begin{equation}
N_{\ell}^{YY} = \left[\sum_{c}  \left(N_{\ell,c}^{YY}\right)^{-1}  \right]^{-1} \ ,
\label{eq:CMBNoiseAddition}
\end{equation}
where $N_{\ell,c}^{YY}$ is the noise on a measurement in a given channel and $Y$ denotes either $T$ or $E$.
Assuming the CMB experiment sees Gaussian beams, this autocorrelation noise is given by
\begin{equation}
N_{\ell,c}^{YY} = \left(\frac{\sigma^Y_c \theta_{\rm FWHM,c}}{T_{\rm
CMB}}\right)e^{\ell(\ell+1)\theta^2_{\rm FWHM,c}/{\rm 8ln2}} \ ,
\label{eq:CMBChannelNoise}
\end{equation}
where $\sigma^Y_c$ is the standard deviation in observable Y (either T
or E) in channel c, and $\theta_{\rm FWHM,c}$ is the FWHM of the
beam, in arc-minutes, at the given frequency channel~\cite{Rocha04}.

For supernovae, the direct observable is their redshift-dependent
magnitude \be m(z) = {\cal M} + 5 \log{d_L} + 25 \label{def:mz} \ee where
${\cal M}$ is the intrinsic supernova magnitude and $d_L$ the luminosity
distance (in Mpc) defined as \be \label{lumdis} d_L(z)=(1+z)
\int_0^z {d{z}' \over H({z}')} \ , \ee where $H(z)$ is the Hubble
parameter with a current value of $H_0$. The information matrix for
SNe observations is \be F^{\mathrm{SN}}_{ab}= \sum_{i}^{N} {1\over
\sigma(z_i)^2 } {\partial m(z_i)\over \partial p_a} {\partial
m(z_i)\over
\partial p_b}. \label{fisher:sne} \ee where the summation is
over the redshift bins and $\sigma_m(z_i)$ is the value given by
Eq.~(\ref{sigma_m}) at the midpoint of the $i$-th bin.

Thus, given a set of theoretical covariance matrices over a given
multipole range, and the specifications for the expected noise in
particular experiments, we can compute the Fisher matrix. The
derivatives with respect to parameters are computed using finite
differences. The values of the finite differences are given in
Appendix \ref{fisher_stepsize}. Upon inverting the
Fisher matrix, we find the theoretical lower limit on the covariance
of the parameters of the model. In this manner, we can forecast how
tightly upcoming experiments will be able to constrain
modified growth.

\subsection{Experiments}
\label{sec:experiments}

The data considered in our forecasts include CMB temperature and polarization (T and E), weak lensing of distant galaxies (WL), galaxy number counts (GC), their
cross-correlations, and SNe observations. We assume CMB T and
E data from the Planck satellite~\cite{Planck}, the galaxy
catalogues and WL data by the \emph{Dark Energy Survey}~\cite{DES} and \emph{Large Synoptic Survey Telescope} (LSST)~\cite{LSST}, complemented by a futuristic SNe data set provided by the ongoing \emph{Canada-France-Hawaii Telescope's} (CFHT) \emph{Supernovae Legacy Survey} (SNLS)~\cite{SNLS,Astier:2005qq}, the \emph{Nearby Supernovae Factory} (NSNF)~\cite{NSNF,aldering02}, and a future \emph{Joint Dark Energy Mission} (JDEM) space mission, such as the \emph{Supernovae/Acceleration Probe} (SNAP)~\cite{SNAP}.

\subsubsection{DES and LSST}

\begin{figure}[t]
\includegraphics[scale=0.2]{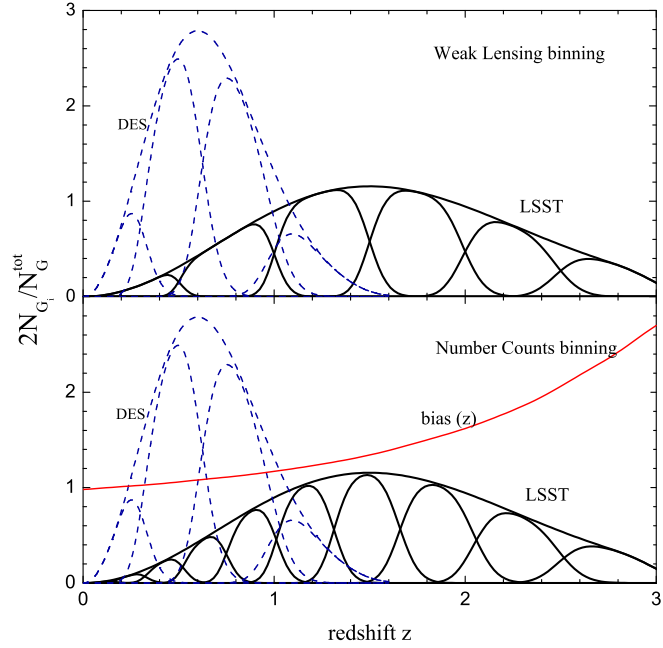}
\caption{The assumed redshift distribution and the partition
into photometric bins for the galaxy counts (GC) and weak lensing
(WL) for DES and LSST. The bias parameter for each GC bin is shown with a thin red solid line
in the lower panel.} \label{Fig:des_lsst_window}
\end{figure}

To describe the expected redshift distribution of galaxies, we take the total galaxy number density to be given by
\begin{equation}
N_G(z) \propto z^2 {\rm exp}(-z/z_0)^{2} \ ,
\end{equation}
which is a a slight modification of the model due to Wittman {\it et al.}~\cite{Wittman}. This function represents a compromise between two trends; more distant galaxies are harder to see, so they are less likely to be a part of the field we observe, but as one goes to higher redshifts, there is more volume of space available and thus one expects there to be more galaxies. The parameter $z_0$ depends on the experiment and defines the redshift at which the most galaxies will be observed. The value of $z_0$ for a given experiment depends on how faint an object it can observe. The window function is suitably normalized so that the total number of galaxies matches that expected from a given survey. The galaxies can be divided into photometric redshift bins, labelled with index $i$,
\begin{equation}
N_G(z) = \sum_i N_{G_i}(z).
\end{equation}
In our analysis, we assume that the photometric
redshift errors are Gaussian distributed,
and that their rms fluctuations increase with redshift as
$\sigma(z)=\sigma_\mathrm{max}(1+z)/(1+z_\mathrm{max})$.
The bin sizes are chosen to increase proportionally to the errors.
The resulting photometric redshift distributions are given by,
\begin{equation}
N_{G_i}(z) = \frac{1}{2} N_G(z)\left[
  \mathrm{erfc}\biggl( \frac{z_{i-1}-z} {\sqrt{2} \sigma(z)}\biggr)
- \mathrm{erfc}\biggl( \frac{z_i-z}{\sqrt{2} \sigma(z)}\biggr) \right],
\label{eq:erfc}
\end{equation}
where erfc is the complementary error function. For a given photometric redshift bin, the normalized selection function that appears in Eq.~(\ref{eq:I_delta}) is given by
\begin{equation} \label{windowfunction}
 W_{G_i}(z) = \frac{N_{G_i}(z)}{N^i} \,
\end{equation}
where $N^i$ is the total number of galaxies in the $i$-th bin.

DES is a project aimed at studying the
nature of the cosmic acceleration, and is planned to start
observations in September, 2009~\cite{DES}. DES includes a 5000
square degree multi-band, optical survey probing the redshift range
$0.1 < z < 1.3$ with a median redshift of $z=0.7$ and an approximate
1-$\sigma$ error of $0.05$ in photometric redshift. In our
simulation, for both WL and galaxy counts, we assume a sky fraction $f_{\rm{sky}}=0.13$, and an angular density of galaxies
$N_G=10$ gal$/$arcmin$^2$. We also assume $\gamma_{\rm rms}=0.18+0.042\,z$, which is the rms shear stemming from the intrinsic
ellipticity of the galaxies and measurement noise, and the photometric redshift uncertainty given by $\sigma(z)=0.05(1+z)$.

LSST is a proposed large
aperture, ground-based, wide field survey telescope~\cite{LSST}. It
is expected to cover up to half of the sky and
catalogue several billion galaxies out to redshift $z\sim 3$. For LSST forecasts, we adopt parameters from the recent review paper by the LSST
collaboration~\cite{Ivezic:2008fe}. Namely, we use $f_{\rm{sky}}=0.5$, $N_{G}=50$ gal$/$arcmin$^2$ for both WL and counts, $\gamma_{\rm
rms}=0.18+0.042\, z$, and $\sigma(z)=0.03\,(1+z)$.

For both DES and LSST, we take the GC photometric bins to be
separated by $5\sigma(z)$. This leads to four redshift bins for DES
and ten for LSST. For WL (source) galaxies, we use four bins for DES
and six for LSST.  In principle, we could have used a larger number
of WL bins for LSST, and it would be interesting to investigate how
the constraints on modified growth improve with finer binning. It is
known, for example, that finer binning does not improve constraints
on the dark energy equation of state $w(z)$~\cite{HuJain}. The
assumed galaxy number distributions for both experiments are shown
in Fig.~\ref{Fig:des_lsst_window}.

The photometric redshift errors used in our analysis should be seen
as the optimistic values for the respective experiments. For a
discussion of potential sources of systematic errors in photo-z
estimates, the reader is referred to~\cite{Huterer:2005ez,Ma:2005rc}. We assume that the systematic errors for both DES and LSST will be comparable or smaller than the statistical errors. While achieving this level of systematics control will be a huge challenge, the white papers for both experiments~\cite{DES,Ivezic:2008fe} describe it as possible and, in fact, necessary for accomplishing the science goals set by these experiments.

\subsubsection{Planck}

The Planck mission~\cite{Planck} of the \emph{European Space Agency} (ESA) is currently expected to launch in the Spring of 2009. Planck will image the
full sky with a sensitivity of $\Delta T/T\sim2\times10^{-6}$, angular
resolution to $5'$, and frequency coverage of $30-857$ GHz~\cite{planckpaper}. The
angular resolution of Planck will be three times superior to
that of NASA's WMAP satellite, and the noise lowered by an order of
magnitude at around $100$ GHz. These significant improvements will permit more accurate measurements of the CMB temperature and polarization power spectra, allowing for a better determination of the cosmological parameters. We use the expected sensitivity and resolution parameters for the lowest three Planck HFI channels based on a 14 month mission~\cite{planckpaper}. We list the parameters relevant to our Fisher analysis in Table~\ref{tab:planck}.
\begin{table}[tb]
\begin{tabular}{r@{~~~~~}r@{~~}r@{~~}r@{~}}
\hline
$\nu$ (GHz)                           & 100  & 143  & 217\\
$\theta_{\mathrm{FWHM}}$ (arc min)   & 10.7 & 8.0  & 5.5\\
$\sigma_T$ ($\mu$K)              & 5.4  & 6.0  & 13.1\\
$\sigma_E$ ($\mu$K)               & n/a  & 11.4 & 26.7\\
$f_\mathrm{sky}$                        & ~    & 0.8 & ~   \\
\hline
\end{tabular}
\caption{ The relevant parameters for Planck~\cite{planckpaper}. We use
the three lowest frequency channels from the HFI.} \label{tab:planck}
\end{table}

\subsubsection{Supernovae and the parameter priors}\label{sec_sn}

\begin{table*}[tb]
\begin{tabular}{@{~}c@{~}l@{     }c@{   }c@{   }c@{   }c@{   }c@{   }c@{   }c@{   }c@{  }
c@{   }c@{   }c@{   }c@{   }c@{   }c@{   }c@{   }c@{   }c@{}}
\hline
& redshift $z$ &~0.1~&~0.2~&~0.3~&~0.4~&~0.5~
&~0.6~&~0.7~&~0.8~&~0.9~&~1.0~&~1.1~&~1.2~&~1.3~&~1.4~&~1.5~&~1.6~&~1.7\\
\hline
& $N(z)$ &~300~&~35~&~64~&~95~&~124~&~150~&~171~&~183~&~179~&~170~&~155~&~142~&~130~&~119~
&~107~&~94~&~80\\
& $\sigma_m(z)/10^{3}$
~&~9~&~25~&~19~&~16~&~14~&~14~&~14~&~14~&~15~&~16~&~17~
&~18~&~20~&~21~&~22~&~23~&~26\\
\hline
\end{tabular}
\caption{The binned redshift distribution of type Ia supernovae used in our analysis. The redshifts given are the upper limits of each bin. Magnitude errors
$\sigma_m(z)$ are evaluated at bin midpoints.}\label{tab:SN}
\end{table*}

In our forecasts, we assume spatially flat geometry and expansion
histories, consistent with the effective equation state of dark
energy equal to $-1$. In addition to the five modified growth
parameters, and the $M$ bias parameters, we vary the Hubble constant
$h$, cold dark matter density $\Omega_ch^2$, the baryon density
$\Omega_bh^2$, the optical depth $\tau$, the scalar spectral index
$n_s$, and the amplitude of scalar perturbations $A_s$. Their
fiducial values are taken to be those from the WMAP 5-year data best
fit~\cite{Dunkley:2008ie}: $\Omega_b h^2 = 0.023, \Omega_c h^2 =
0.11, h=0.72, \tau = 0.087, n_s=0.963$. The fiducial values for
bias parameters are motivated by the parametrized halo model
described in \cite{HuJain}, and we show them in the lower panel of
Fig.~\ref{Fig:des_lsst_window}. Imposing a prior on the value of
$h$ from the Hubble Space Telescope (HST) did not make a noticeable
difference in our results.

In addition, to better constrain the background expansion parameters, we include data from a futuristic set of SNe luminosity distances from SNLS, NSNF and SNAP. SNe observations determine the magnitude of the SNe, $m(z)$, defined in Eq.~(\ref{def:mz}).
The uncertainty in $m(z)$, at any $z$-bin containing $N_{\mathrm{bin}}$ supernovae, is given by
\begin{equation}
\sigma_m(z) = \sqrt{\frac{\sigma_\mathrm{obs}^2}{N_{\mathrm{bin}}}+ \mathrm d m^2},
\label{sigma_m}
\end{equation}
and we assume $\sigma_\mathrm{obs} = 0.15$.
The systematic error, $\mathrm d m$, is assumed to increase linearly with redshift:
\begin{equation}
\mathrm d m = \delta m \frac{z}{z_{\mathrm{max}}},
\end{equation}
$\delta m$ being the expected uncertainty and $z_{\mathrm{max}}$ the maximum
redshift. In our analysis, we follow~\cite{Kim,Pogosian:2005ez}
and assume $\delta m =  0.02$ and $z_{\mathrm{max}} = 1.7$ for SNe from SNAP, plus low-$z$ SNe from NSNF. We list the relevant parameters in Table \ref{tab:SN}. The absolute magnitude, or the so-called nuisance parameter ${\cal M}$, is treated as an undetermined parameter in our analysis.


\section{Results}
\label{results}

\begin{figure}[tb]
\includegraphics[scale=0.2]{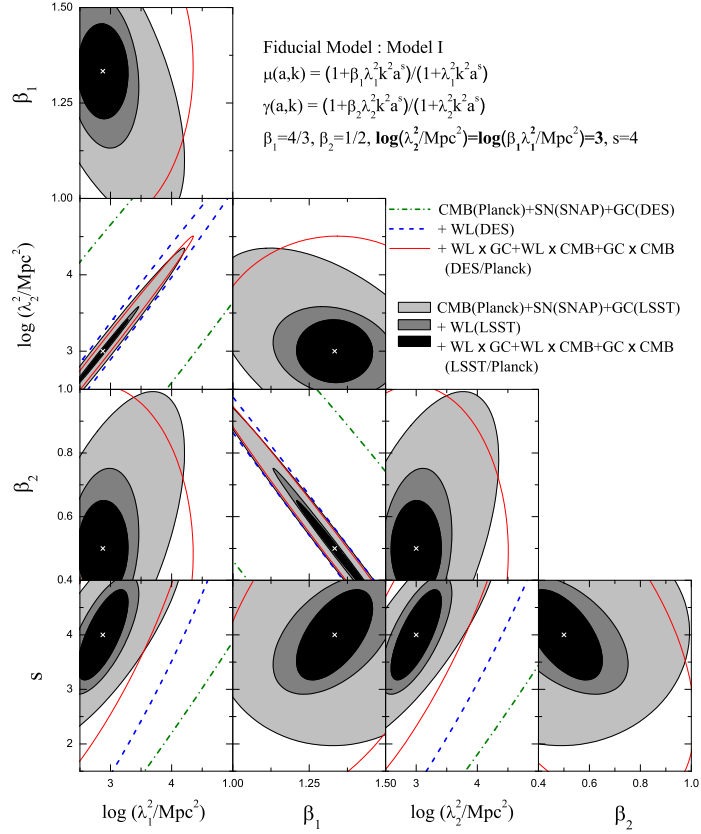}
\caption{The $68$\% confidence contours for our five modified growth (MG) parameters for Model I, based on f(R) (described in Sec.~\ref{theory}), as constrained by several different combinations of correlation functions from various experiments (shown in the legend). As described in detail in Sec.~\ref{theory}, $\lambda^2_i$ represent length scales, $\beta_i$ represent couplings and $s$ encodes the time evolution of the characteristic mass scale of the model.} \label{fig:model1}
\end{figure}

\begin{figure}[tb]
\includegraphics[scale=0.2]{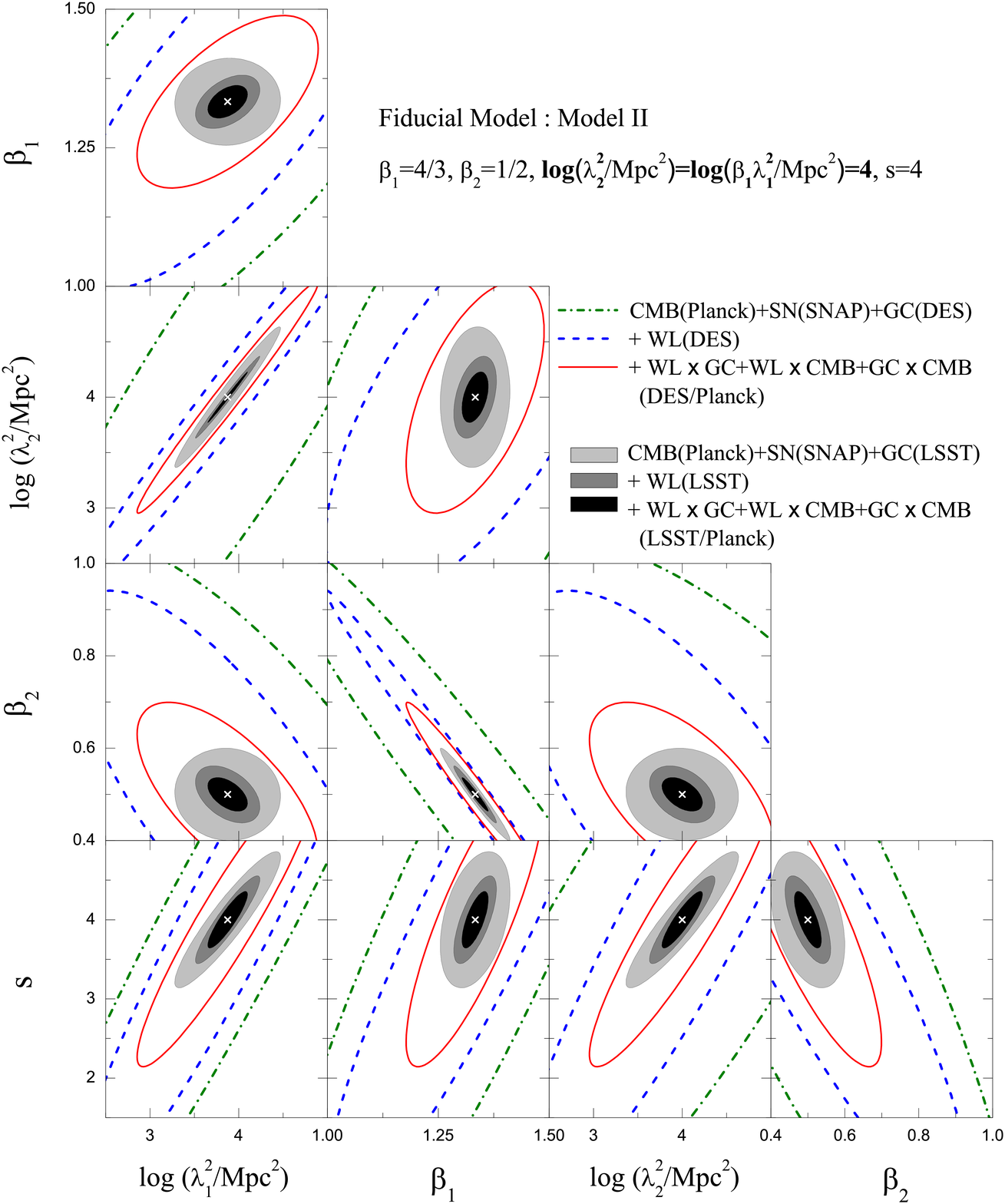}
\caption{The $68$\% confidence contours for the five MG parameters for Model II, based on $f(R)$ with a different Compton scale (described in Sec.~\ref{theory}), as constrained by several different combinations of correlation functions from various experiments (shown in the legend).} \label{fig:model2}
\end{figure}

\begin{figure}[tb]
\includegraphics[scale=0.2]{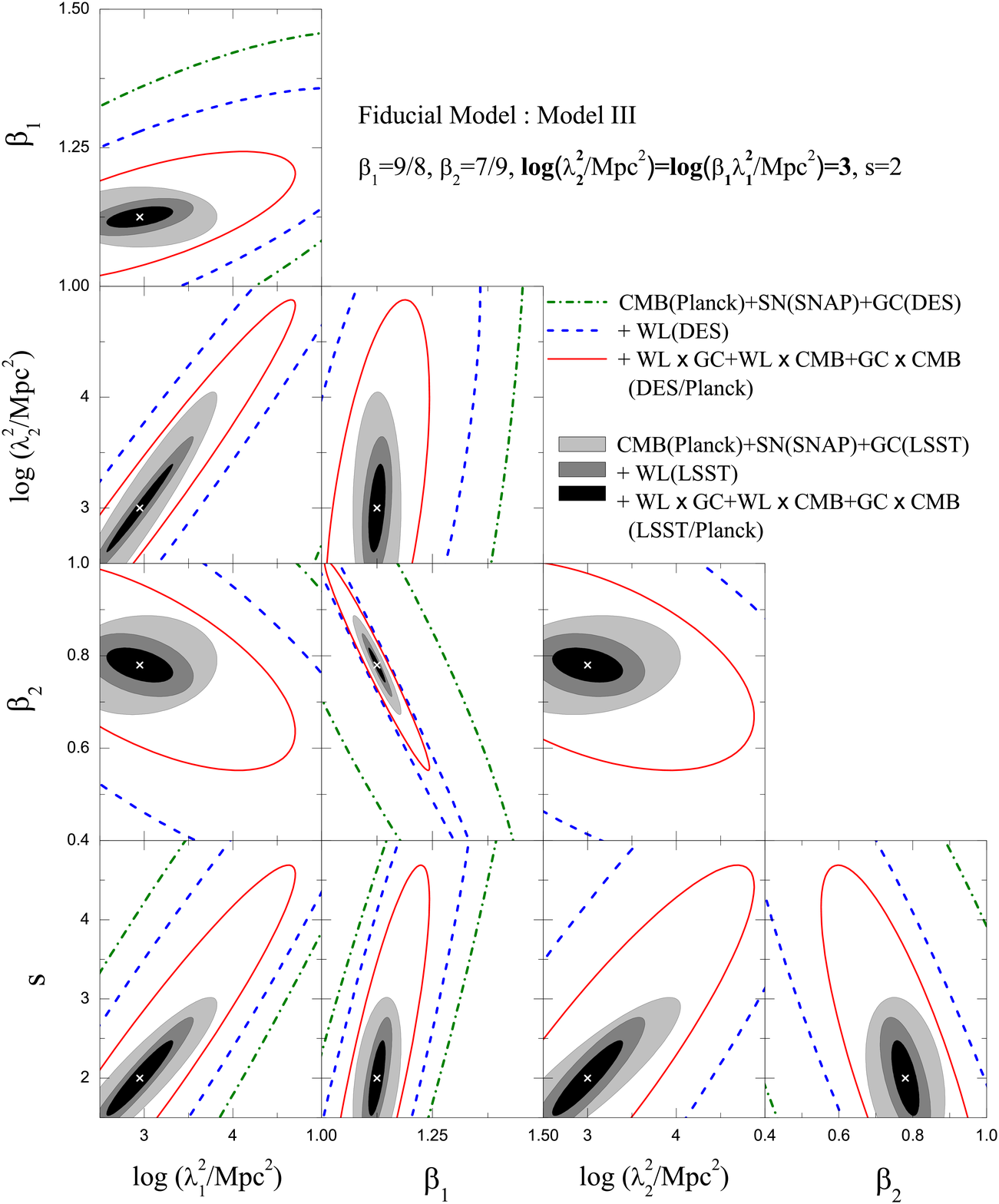}
\caption{The $68$\% confidence contours for the five MG parameters for Model III, based on a Chameleon model (described in Sec.~\ref{theory}), as constrained by several different combinations of correlation functions from various experiments (shown in the legend).} \label{fig:model3}
\end{figure}

\begin{figure}[tb]
\includegraphics[scale=0.2]{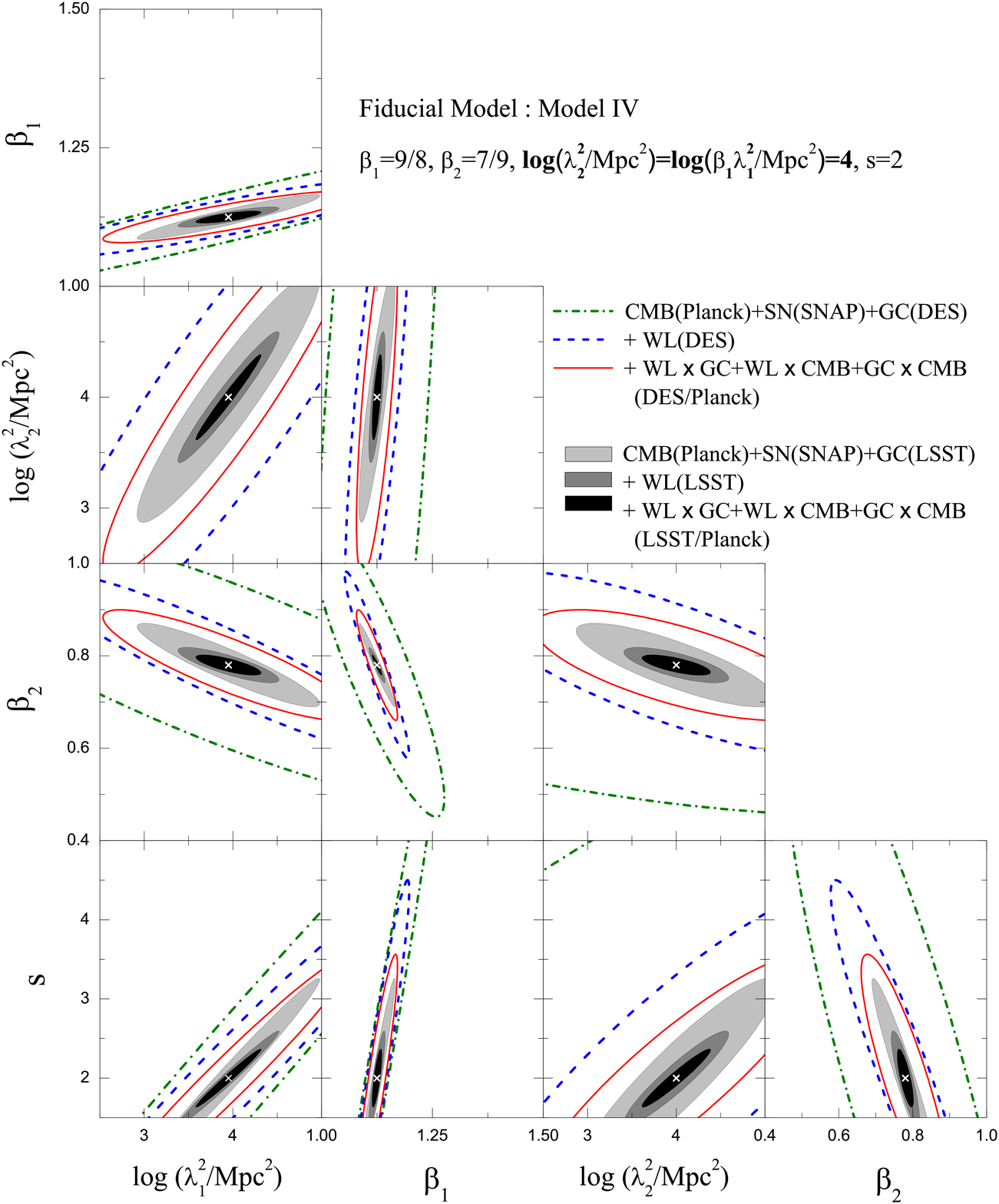}
\caption{The $68$\% confidence contours for the five MG parameters for Model IV, based on a Chameleon model with a different Compton scale (described in Sec.~\ref{theory}), as constrained by several different combinations of correlation functions from various experiments (shown in the legend).} \label{fig:model4}
\end{figure}


\begin{table*}[tb]
\begin{tabular}{|c|cc|ccccc|ccccc|ccccc|ccccc|}
\hline
\multicolumn{3}{|c|}{} & \multicolumn{20}{c|}{$1\sigma$ errors  from \textbf{Planck}, \textbf{SNAP} and \textbf{DES}}\\
 & & & \multicolumn{5}{c|}{GC+CMB+SN} & \multicolumn{5}{c|}{+WL}&\multicolumn{5}{c|}{+WL$\times$GC}&\multicolumn{5}{c|}{ALL}
   \\
Model&\textbf{P}&${\rm fiducial}$    & &A  & &   &B & &A   &    &B  &   &  &A  &   &B &  &  &A   &   &B &\\
\hline \hline

&$\lama$        &  $3+{\rm log}(3/4)$ &  &3.8   &&  &1.9    &  &1.7    &  &1.1   &&  &0.99  &   &0.71  &&  &0.99  &  &0.71&\\
&$\beta_{1}$    &       $4/3$         &  &0.71  &&  &0.55   &  &0.46   &  &0.34  &&  &0.27  &   &0.22  &&  &0.27  &  &0.22&\\
I&$\lamb$       &        $3$          &  &3.8   &&  &2.1    &  &1.9    &  &1.2   &&  &1.0   &   &0.75  &&  &1.0   &  &0.74&\\
&$\beta_{2}$    &       $1/2$         &  &0.85  &&  &0.70   &  &0.59   &  &0.45  &&  &0.34  &   &0.29  &&  &0.34  &  &0.29&\\
&$s$            &        $4$          &  &6.1   &&  &2.6    &  &3.7    &  &1.9   &&  &2.1   &   &1.4   &&  &2.1   &  &1.4&\\
\hline
&$\lama$        &  $4+{\rm log}(3/4)$ &  &1.7   &&  &0.58   &  &1.2    &  &0.46  &&  &0.68  &   &0.35  &&  &0.68  &  &0.35& \\
&$\beta_{1}$    &       $4/3$         &  &0.31  &&  &0.09   &  &0.22   &  &0.08  &&  &0.10  &   &0.06  &&  &0.10  &  &0.06& \\
II&$\lamb$      &        $4$          &  &2.1   &&  &1.2    &  &1.2    &  &0.59  &&  &0.70  &   &0.41  &&  &0.70  &  &0.41& \\
&$\beta_{2}$    &       $1/2$         &  &0.38  &&  &0.14   &  &0.29   &  &0.11  &&  &0.13  &   &0.08  &&  &0.13  &  &0.08& \\
&$s$            &        $4$          &  &3.5   &&  &0.95   &  &2.5    &  &0.83  &&  &1.2   &   &0.67  &&  &1.2   &  &0.66& \\

\hline
&$\lama$        &  $3+{\rm log}(8/9)$ &  &2.3   &&  &0.83   &  &1.9    &  &0.76  &&  &1.2   &   &0.62  &&  &1.2   &  &0.62&\\
&$\beta_{1}$    &        $9/8$        &  &0.22  &&  &0.08   &  &0.16   &  &0.06  &&  &0.08  &   &0.05  &&  &0.08  &  &0.05&\\
III&$\lamb$     &         $3$         &  &3.5   &&  &2.5    &  &2.1    &  &1.0   &&  &1.3   &   &0.78  &&  &1.3   &  &0.78&\\
&$\beta_{2}$    &        $7/9$        &  &0.42  &&  &0.23   &  &0.31   &  &0.14  &&  &0.15  &   &0.10  &&  &0.15  &  &0.10&\\
&$s$            &         $2$         &  &4.2   &&  &1.2    &  &3.3    &  &1.2   &&  &1.8   &   &0.95  &&  &1.8   &  &0.95&\\
\hline
&$\lama$        &  $4+{\rm log}(8/9)$ &  &2.5   &&  &0.75   &  &1.4    &  &0.66  &&  &0.95  &   &0.53  &&  &0.94  &  &0.53&\\
&$\beta_{1}$    &        $9/8$        &  &0.10  &&  &0.03   &  &0.05   &  &0.03  &&  &0.03  &   &0.02  &&  &0.03  &  &0.02&\\
IV&$\lamb$      &         $4$         &  &3.9   &&  &3.0    &  &1.6    &  &0.99  &&  &1.1   &   &0.77  &&  &1.1   &  &0.76&\\
&$\beta_{2}$    &        $7/9$        &  &0.22  &&  &0.11   &  &0.13   &  &0.06  &&  &0.08  &   &0.04  &&  &0.08  &  &0.04&\\
&$s$            &         $2$         &  &3.2   &&  &1.1    &  &1.7    &  &0.89  &&  &1.0   &   &0.68  &&  &1.0   &  &0.68&\\
\hline\hline
\end{tabular}
\caption{Error forecasts for parameters describing the
modification of gravity based on the fiducial models considered in the paper, when DES is used for the GC and WL measurements. We list results for combinations of weak lensing (WL), galaxy number
counts (GC), cross-correlations between lensing and
counts(WL$\times$GC), lensing and CMB (WL$\times$CMB), counts and
CMB (GC$\times$CMB) and the combination of all of these (ALL). SNAP and Planck are always included in order to constrain the background parameters. We
also illustrate the effect of adding a prior on bias parameters in columns A (no bias
priors) and B ($\sigma$(bias)=0.001). } \label{tab:DES}
\end{table*}

\begin{table*}[tb]
\begin{tabular}{|c|cc|ccccc|ccccc|ccccc|ccccc|}
\hline
\multicolumn{3}{|c|}{} & \multicolumn{20}{c|}{$1\sigma$ errors  from \textbf{Planck}, \textbf{SNAP} and \textbf{LSST}}\\
 & & & \multicolumn{5}{c|}{GC+CMB+SN} & \multicolumn{5}{c|}{+WL}&\multicolumn{5}{c|}{+WL$\times$GC}&\multicolumn{5}{c|}{ALL}
   \\
Model&\textbf{P}&${\rm fiducial}$    & &A  & &   &B & &A   &    &B  &   &  &A  &   &B &  &  &A   &   &B &\\
\hline
\hline

&$\lama$        &  $3+{\rm log}(3/4)$ &  &0.89  &&  &0.64   &  &0.39   &  &0.27  &&  &0.27  &   &0.18  &&  &0.27  &  &0.18&\\
&$\beta_{1}$    &       $4/3$         &  &0.27  &&  &0.20   &  &0.13   &  &0.11  &&  &0.08  &   &0.07  &&  &0.08  &  &0.07&\\
I&$\lamb$       &        $3$          &  &0.90  &&  &0.65   &  &0.39   &  &0.28  &&  &0.27  &   &0.19  &&  &0.27  &  &0.19&\\
&$\beta_{2}$    &       $1/2$         &  &0.33  &&  &0.25   &  &0.17   &  &0.14  &&  &0.10  &   &0.09  &&  &0.10  &  &0.09&\\
&$s$            &        $4$          &  &1.35  &&  &0.94   &  &0.73   &  &0.54  &&  &0.55  &   &0.45  &&  &0.55  &  &0.45&\\
\hline
&$\lama$        &  $4+{\rm log}(3/4)$ &  &0.40  &&  &0.21   &  &0.24   &  &0.12  &&  &0.15  &   &0.09  &&  &0.15  &  &0.09& \\
&$\beta_{1}$    &       $4/3$         &  &0.05  &&  &0.03   &  &0.03   &  &0.02  &&  &0.02  &   &0.01  &&  &0.02  &  &0.01& \\
II&$\lamb$      &        $4$          &  &0.42  &&  &0.24   &  &0.25   &  &0.14  &&  &0.15  &   &0.09  &&  &0.15  &  &0.09& \\
&$\beta_{2}$    &       $1/2$         &  &0.07  &&  &0.05   &  &0.04   &  &0.03  &&  &0.02  &   &0.02  &&  &0.02  &  &0.02& \\
&$s$            &        $4$          &  &0.57  &&  &0.32   &  &0.37   &  &0.22  &&  &0.24  &   &0.18  &&  &0.24  &  &0.18& \\

\hline
&$\lama$        &  $3+{\rm log}(8/9)$ &  &0.58  &&  &0.29   &  &0.40   &  &0.19  &&  &0.25  &   &0.14  &&  &0.25  &  &0.14&\\
&$\beta_{1}$    &        $9/8$        &  &0.04  &&  &0.03   &  &0.02   &  &0.02  &&  &0.01  &   &0.009 &&  &0.01  &  &0.009&\\
III&$\lamb$     &         $3$         &  &0.70  &&  &0.42   &  &0.43   &  &0.25  &&  &0.27  &   &0.16  &&  &0.27  &  &0.16&\\
&$\beta_{2}$    &        $7/9$        &  &0.07  &&  &0.06   &  &0.05   &  &0.04  &&  &0.02  &   &0.02  &&  &0.02  &  &0.02&\\
&$s$            &         $2$         &  &0.68  &&  &0.31   &  &0.51   &  &0.24  &&  &0.32  &   &0.20  &&  &0.32  &  &0.20&\\
\hline
&$\lama$        &  $4+{\rm log}(8/9)$ &  &0.69  &&  &0.20   &  &0.38   &  &0.17  &&  &0.25  &   &0.14  &&  &0.25  &  &0.14&\\
&$\beta_{1}$    &        $9/8$        &  &0.03  &&  &0.007  &  &0.01   &  &0.004 &&  &0.007 &   &0.003 &&  &0.007 &  &0.003&\\
IV&$\lamb$      &         $4$         &  &0.75  &&  &0.35   &  &0.39   &  &0.21  &&  &0.26  &   &0.15  &&  &0.26  &  &0.15&\\
&$\beta_{2}$    &        $7/9$        &  &0.06  &&  &0.02   &  &0.03   &  &0.01  &&  &0.01  &   &0.007 &&  &0.01  &  &0.007&\\
&$s$            &         $2$         &  &0.84  &&  &0.25   &  &0.40   &  &0.18  &&  &0.25  &   &0.15  &&  &0.25  &  &0.15&\\
\hline\hline
\end{tabular}
\caption{Error forecast for the parameters describing the
modification of gravity based on the fiducial models considered in Sec.~\ref{theory}, when LSST is used for the GC and WL measurements. We list the results for combinations of weak lensing (WL), galaxy number
counts (GC), cross-correlations between lensing and
counts(WL$\times$GC), lensing and CMB (WL$\times$CMB), counts and
CMB (GC$\times$CMB) and the combination of all of these (ALL). SNAP and Planck are always included in order to constrain the background parameters. We
also illustrate the effect of adding a prior on the bias parameters in columns A (no bias
priors) and B ($\sigma$(bias)=0.001). } \label{tab:LSST}
\end{table*}

\begin{figure}[tb]
\includegraphics[scale=0.2]{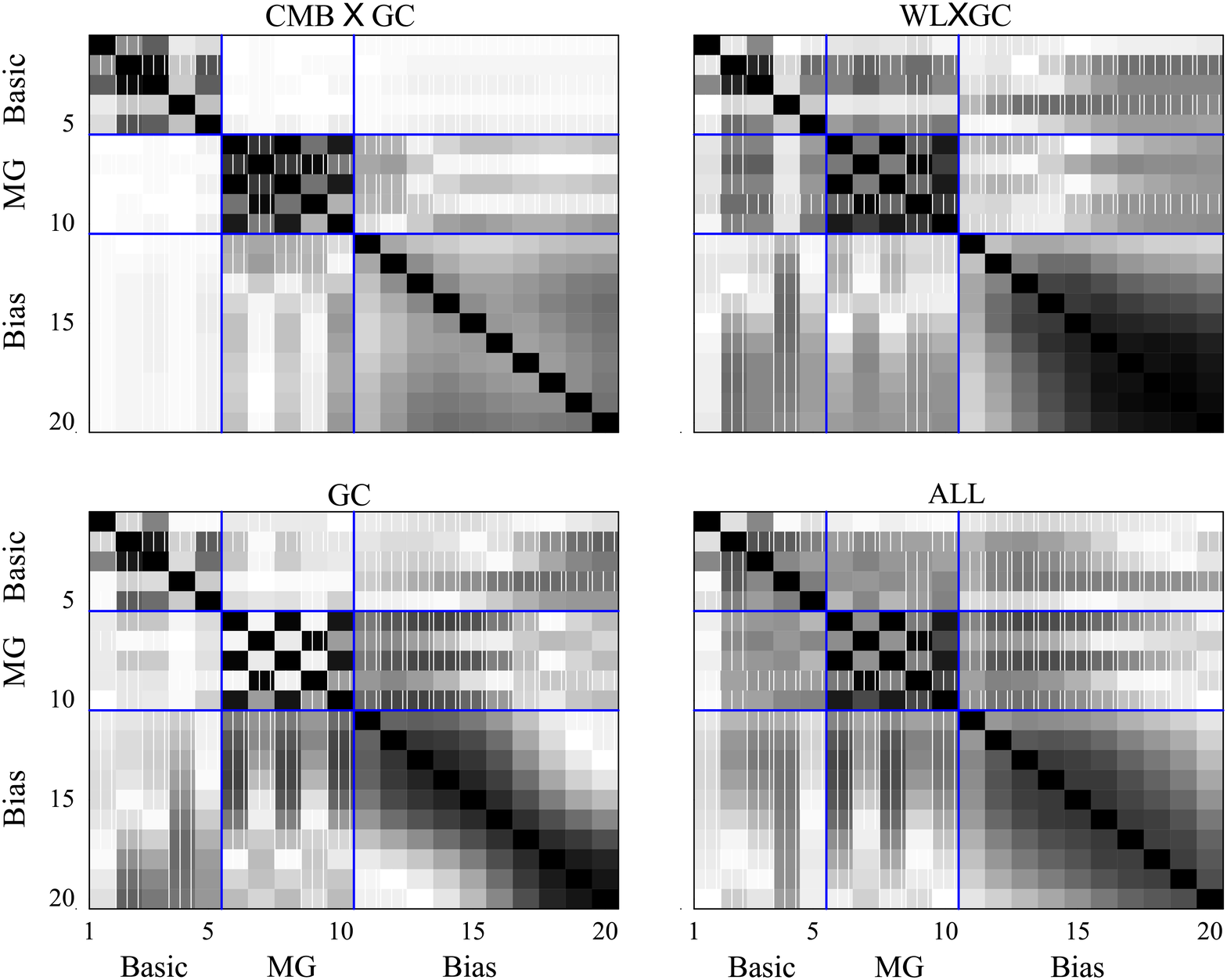}
\includegraphics[scale=0.2]{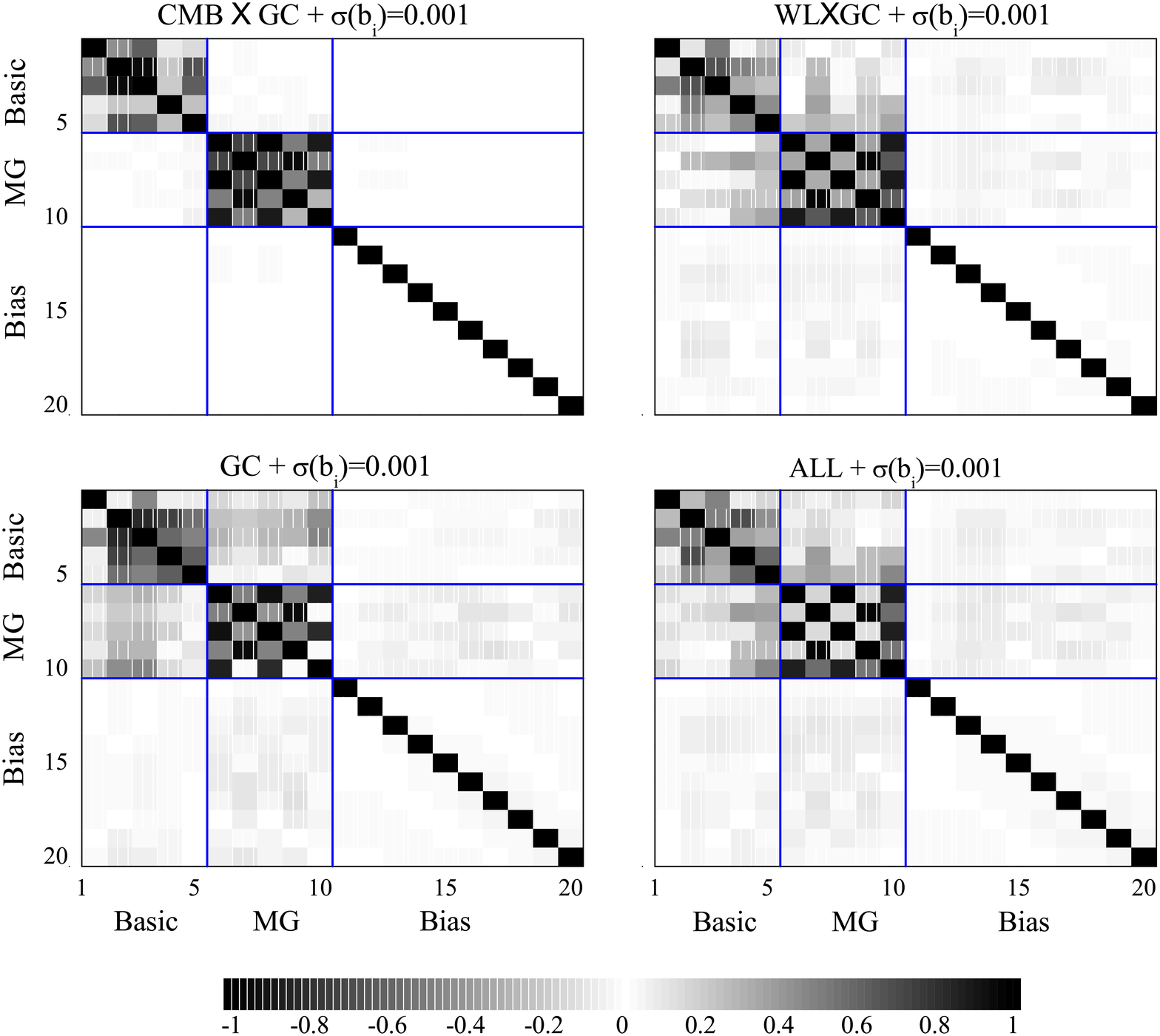}
\caption{Correlation matrices for the basic, MG, and the bias
parameters for the $f(R)$ fiducial model with $\lamb=4$ (Model II).
Parameters $1$ to $5$ are the basic cosmological parameters, namely,
$\Omega_bh^{2}$, $\Omega_ch^{2}$, $h$, $\tau$ and $n_s$
respectively; parameters $6$ to $10$ are the MG parameters, namely,
$\lama,\beta_1,\lamb, \beta_2$ and $s$ respectively. Parameters $11$
to $20$ denote the $10$ bias parameters for the LSST GC bins from
low to high redshift. The positive and negative correlations are
shown with shaded blocks with and without vertical lines. We compare
the correlation matrices with(lower panel) and without(upper panel)
imposing a bias prior of $\sigma(b_i)=0.001$.} \label{fig:corr}
\end{figure}

\begin{table*}[tb]
\begin{tabular}{c|cc|cc|cc|cc}
\hline\hline   & \multicolumn{2}{c|}{Model
I}&\multicolumn{2}{c|}{Model II}&\multicolumn{2}{c|}{Model
III} & \multicolumn{2}{c}{Model IV}
   \\
{\textbf{P}} &DES       &LSST       &DES          &LSST       &DES        &LSST       &DES        &LSST\\
\hline
$\lama$      &$34\%$    &$9\%$       &$18\%$      &$4\%$      &$41\%$     &$8\%$      &$24\%$     &$6\%$\\
$\beta_1$    &$20\%$    &$6\%$       &$9\%$       &$2\%$      &$7\%$      &$0.9\%$    &$3\%$      &$0.6\%$\\
$\lamb$      &$33\%$    &$9\%$       &$18\%$      &$4\%$      &$43\%$     &$9\%$      &$28\%$     &$7\%$\\
$\beta_2$    &$68\%$    &$20\%$      &$26\%$      &$4\%$      &$19\%$     &$3\%$      &$10\%$     &$1\%$\\
$s$          &$53\%$    &$14\%$      &$30\%$      &$6\%$      &$90\%$     &$16\%$     &$50\%$     &$13\%$\\
 \hline
 \hline
\end{tabular}
\caption{Relative errors on the five modified growth parameters
from all data combined. This table allows for a quick comparison between DES and LSST, and between different fiducial models. For more complete
information see Tables~\ref{tab:DES} and \ref{tab:LSST}.} \label{tab:final}
\end{table*}

\begin{figure}[tb]
\includegraphics[scale=0.2]{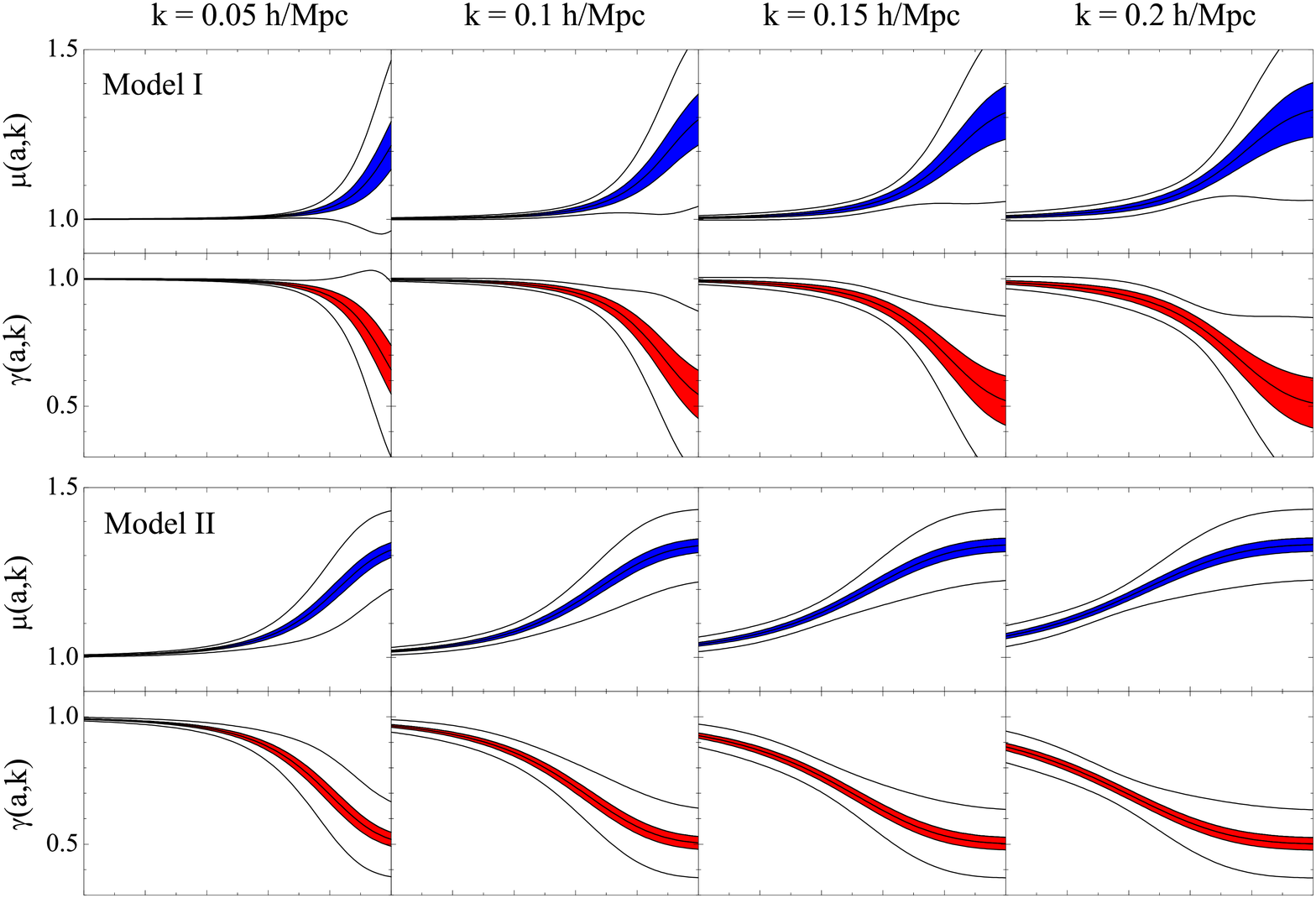}
\includegraphics[scale=0.2]{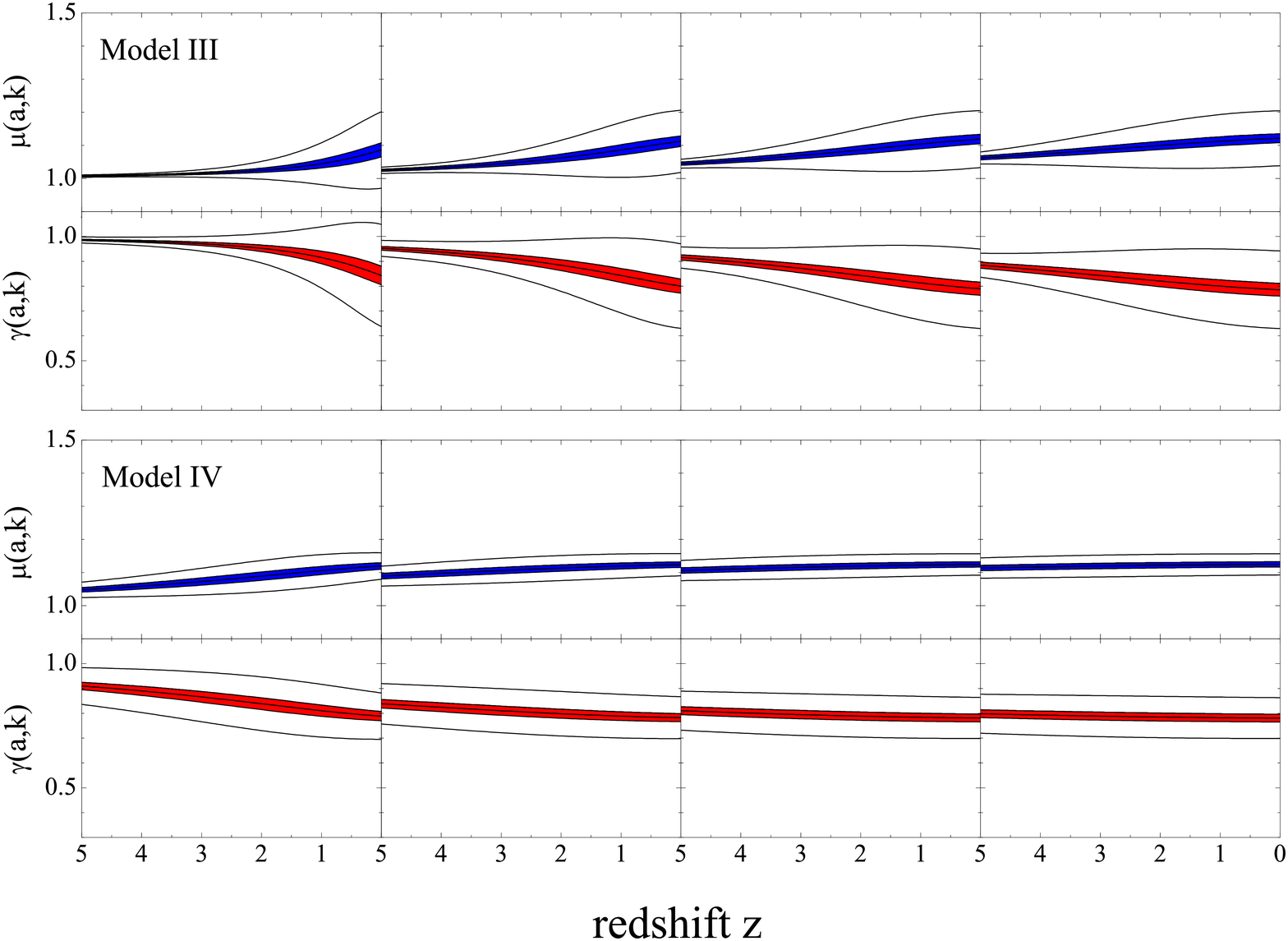}
\caption{Reconstruction of the functions $\mu(a,k)$ and $\gamma(a,k)$ from the
constraints on the parameters $\{\lambda_i^2,\beta_i,s\}$, for Models I-IV. The
inner (color-shaded) areas show the $1\sigma$ errors from
LSST+Planck+SNAP, while the outer contours delimited by a solid line show the $1 \sigma$ errors from  DES+Planck+SNAP. The line in the
center denotes the fiducial model.} \label{fig:reconst_fr}
\end{figure}

Using the methods detailed in Section \ref{observables}, we have
evaluated the Fisher errors on the \emph{Modified Growth} (MG)
parameters \{$s,\beta_1,\beta_2,{\rm log}(\lambda_1^2/{\rm
Mpc}^2),{\rm log}(\lambda_2^2/{\rm Mpc}^2)$\}, as well as the usual
set of standard cosmological parameters. In
Figs.~\ref{fig:model1}-\ref{fig:model4} we plot the $68$\% C.L.
contours for the MG parameters. Each of the ellipses is plotted
after marginalizing over all other parameters that were varied. The
plots show contributions of different types of data: GC, GC+WL,
and the combination of all data. We represent the contours from DES
using different types of lines, while contours from LSST are shown using
different shades.

We find that the five parameters are generally
correlated and the degeneracy between $\lama$ and $\lamb$ is
strongly positive, while $\beta_1$ and $\beta_2$ have a strong negative
correlation. The negative correlation between $\beta$'s is expected. One
can enhance the growth by either raising $\beta_1$, which increases the effective
Newton's constant, or lowering $\beta_2$, which enhances the relative strength
of the Newtonian potential $\Psi$ which drives the clustering of matter (see Eq.~(\ref{density_evol})). This also explains the degeneracy between the
two length scale parameters $\lama$ and $\lamb$.

The full list of marginalized Fisher errors for the four fiducial
models are listed in Table~\ref{tab:DES} for DES and
Table~\ref{tab:LSST} for LSST. Comparing the constraints on the MG
parameters from different datasets, we find that the
constraining powers of WL, GC and WL$\times$GC are comparable and
that they provide much more stringent constraints than WL$\times$CMB
and GC$\times$CMB. This is, in part, because there are many more
different correlations between WL and GC than there are
cross-correlations with CMB (see Fig.~\ref{Fig:table}). Also,
correlations of GC and WL with CMB suffer from a larger statistical
uncertainty, since only the ISW part of the total CMB anisotropy is
correlated with large scale structure.

As expected, models with $\lamb=4$ (II, IV) are better constrained than
models with $\lamb=3$ (I, III). In the former case,
variations of the MG parameters give rise to larger changes in $\mu$
and $\gamma$ on linear scales. Also, the $\beta_i$ parameters, which are directly related to the coupling $\alpha'$ in scalar-tensor theories, are much better constrained in Models III and IV, compared to Models I,II. This is because with $s=2$, functions $\mu$ and $\gamma$ depart from
their GR values of unity at much earlier times for the same scales (as can be seen in Fig.~\ref{fig:mu_gamma}). Hence, with DES, and especially with LSST, one will be able to
significantly reduce the volume of the allowed parameter space in Chameleon type models.
Also, for all models we have considered, the Fisher errors are small enough to allow for
meaningful consistency checks using Eq.~(\ref{st_consistency}).

We can quantify the correlation between two randomly distributed
parameters $p,q$ by their correlation coefficient, which is defined
as
\be\label{cc} \xi(p,q)\equiv\frac{{\rm
Cov}(p,q)}{\sigma_{p}\sigma_{q}} \ ,
\ee
where ${\rm Cov}(p,q)$ and
$\sigma_p,\sigma_q$ are the off-diagonal elements of the covariance
matrix and the variance of the parameters $p,q$, respectively. For example, with all data
combined, $\xi[\,\lama,\lamb\,]\sim0.9$ and $\xi(\beta_1,
\beta_2)\sim-0.9$ for both models I and II. This strong
degeneracy appears in both DES and LSST data, which reflects the
degeneracy between $\mu$ and $\gamma$.

The scale-dependent growth pattern induced by modified gravity can,
in principle, be partially degenerate with the bias factors of
galaxies. Thus, in Fig.~\ref{fig:corr}, we plot the correlation
coefficients of the basic cosmological parameters (lower-left
blocks), MG parameters (central blocks) and the bias factors
(upper-right blocks) using model II and LSST as an example. In each
panel, the off-diagonal blocks show the correlation among the basic,
MG and bias parameters. In the case of GC$\times$CMB, the MG
parameters are correlated with bias for all the redshift bins, since
GC$\times$CMB is affected by modifications in our fiducial model
over the entire redshift range of the LSST GC bins (see
Fig.~\ref{fig:kernel}). Hence, the signal seen by GC$\times$CMB can
be either from modified gravity or from galaxy bias. For GC,
the MG parameters are more strongly correlated with biases for
redshift bins centered at $z<2$ than for the higher redshift bins,
because GC cannot detect the MG effects from our fiducial model at $z>2$, as
illustrated in Fig.~\ref{fig:kernel}. There is less degeneracy in
the WL$\times$GC panel than in the GC panel, since WL is
independent of bias. Na\"{i}vely, one might expect much less
degeneracy among bias and MG parameters after combining all of the data,
including WL. However, since WL data contains degeneracies among basic
parameters and MG parameters, the degeneracy cannot be effectively
broken. This is what we see in the upper panel of
Fig.~\ref{fig:corr}.

It is interesting to consider the case where we somehow have
measured the bias independently. In Tables~\ref{tab:DES} and
\ref{tab:LSST}, we show the results with and without adding a strong
prior on the bias parameters: $\sigma(b_{i})=0.001$. The results
are listed in the A and B columns in Tables~\ref{tab:DES}
and~\ref{tab:LSST}, and the correlation matrix is shown in the
lower panel of Fig.~\ref{fig:corr}. For the case considered in Fig.~\ref{fig:corr},
the bias prior can help to improve the constraints on MG parameters, on average, by
$19$\%, $42$\%, $30$\% and $31$\% for GC$\times$CMB, GC,
WL$\times$GC and all data combined, respectively. Comparing these
numbers with Fig.~\ref{fig:corr}, one can see that the stronger
the degeneracy between the MG parameters and the bias factors, the
bigger improvement we can obtain by breaking the degeneracy with a
bias prior. Moreover, thanks to this strong bias prior, the
relic degeneracy among basic parameters, MG parameters and bias is
effectively broken as shown in Fig.~\ref{fig:corr}.

We note the importance of not ignoring correlations between galaxy counts from neighboring bins. Including the galaxy cross-spectra is necessary for the estimation of covariances from the Fisher matrix (Eq.~(\ref{eq:Fisher})), since bins do have an overlap. Its inclusion also helps to determine the bias and, hence, improve the constraints on the MG parameters. The cross-correlations between widely separated bins vanish and only the neighboring bins actually contribute. Still, without their inclusion, we find that the costraints on the MG parameters from LSST are weakened by up to 40\% in the case with no prior on the bias. Adding a strong prior on the bias nullifies the relative importance of the infomation contribution from galaxy cross-spectra.

Even without the bias prior, we did not find significant
correlations between the MG parameters and the basic cosmological
parameters, such as $\Omega_{b}h^{2}, \Omega_{c}h^{2},\etc$, namely,
the absolute value of the corresponding correlation coefficients are less than $0.3$.

Comparing the constraints from DES with LSST, we find that,
typically, LSST can improve constraints on individual MG parameters by a factor of $3$ or better.
With all tomographic data combined (no priors used), the relative
errors on MG parameters, which is the $1\sigma$ marginalized
error divided by the corresponding fiducial values, are summarized
in Table~\ref{tab:final}. There we can see that even DES will have
significant power to constrain some of the MG parameters, giving $20\%-40$\% level
constraints, while LSST can go below $10$\% levels.

Given the constraints on MG parameters, one can reconstruct $\mu$
and $\gamma$ using error propagation. The results are shown in
Fig.~\ref{fig:reconst_fr} for Models I-IV.  The inner shaded bands are the $68$\% C.L. regions from all combined data from LSST, and the outer
lines show the constraints from DES. We see that even DES is able to
put good constraints on $\mu, \gamma$ and test our fiducial
models at high confidence level. In the plot, we see that the errors are
generally smaller at small $k$'s and high redshifts. However, this
does not really mean that we can constrain $\mu$ and $\gamma$ very
precisely in that region. Instead, this is likely an artifact of our
parametrization. One generally expects to find ``sweet spots'', i.e. regions where the errors are small, close to the transition scale. This is analogous to the case of the
dark energy equation-of-state $w(z)$, where it is know that the number and the locations of
``sweet spots'' strongly depend on the parametrization of $w(z)$~\cite{Linder:2005ne}.

\section{Conclusions}\label{summary}

In this paper, we have investigated the power of future
tomographic surveys to constrain modifications in the growth of structure w.r.t. to that predicted in $\Lambda$CDM. Models of modified gravity, as well as models of coupled dark energy and dark matter, in general introduce a scale-dependence in the growth of structure and a time- and scale-dependent slip between the gravitational potentials. These modifications are expected to leave characteristic imprints on the observables, which in principle could be used to break the background degeneracy among different models of cosmic acceleration. It is therefore useful and important to explore to which extent the upcoming experiments will be able to detect and constrain modified growth patterns.

We have used a five-parameter description for the rescaling of the Newton constant, $\mu(a,k)$, and for the ratio of the metric potentials, $\gamma(a,k)$, equivalent to that introduced in~\cite{Bertschinger:2008zb}. From the point of view of scalar-tensor theories (e.g.$f(R)$ and Chameleon models) these parameters are  related to the coupling in the dark sector and to the characteristic mass scale of the model. In these cases, the five parameters are not all independent; they need to satisfy the consistency conditions given by Eq.~(\ref{st_consistency}).  This can be used to constrain the scalar-tensor models and potentially rule them out. In Sec.~\ref{theory} we have described the $f(R)$ and Chameleon theories which we have used as fiducial models for our error forecasting. We have then studied in detail the constraints on the five parameters based on four fiducial models (two $f(R)$ and two Chameleon) expected from Weak Lensing (WL), Galaxy Counts (GC), CMB and their cross-correlation spectra as seen by Planck, DES and
LSST (additionally using Planck and SNAP to constrain the standard set of cosmological parameters). We have found that for scalar-tensor type models, DES can provide $20-40$\%
level constraints on individual parameters, and that LSST can improve on that, constraining the MG parameters to better than $10$\%
level.  We have also found a strong degeneracy between the two functions $\mu$ and $\gamma$, which was consistent with our expectations. Despite this degeneracy, however, the error ellipses are sufficiently tight for us to still find non-trivial bounds on the parameters of both functions. Overall, with DES, and especially with LSST, one will be able to
significantly reduce the volume of the allowed parameter space in scalar-tensor type models.

We have also found that
the dilution of the constraints due to the degeneracy with linear
bias is minimal when the full set of auto- and cross-correlation
tomographic datasets is considered.


The method we have employed  relies on the choice of fiducial values for the MG parameters, and therefore is model-dependent. Nevertheless, the results we have obtained are encouraging as they show that upcoming and future surveys can place non-trivial bounds on modifications of the growth of structure  even in the most conservative case, i.e. considering only linear scales. This motivates us to apply the Principal Component Analysis (PCA) to the functions $\mu$ and $\gamma$. This method is more demanding computationally but, on the other hand, allows for a model-independent comparison of  different experiments according to the relative gain in information about the functions. Furthermore, this method can point to the ``sweet spots'' in redshift and scale where data is most sensitive to variations in the functions $\mu$ and $\gamma$. This would be important information for designing future observing strategies. We present the results of the PCA analysis in a separate publication~\cite{modgrav_pca}.

Another direction for future work is to study the effects of allowing for a small fraction of hot dark matter, such as neutrinos, and dynamical dark energy. One can, in principle, model the effects of dark energy perturbations and massive neutrinos by appropriately choosing the time-scale dependence of $\mu$ and $\gamma$. It is generally difficult to design a test that will definitively distinguish modified gravity from an exotic form of dark energy. However, it is clear that future data will have the potential to detect or, at the very least, significantly reduce the range of, departures from the $\Lambda$CDM model.


\acknowledgments We would like to thank Rachel Bean, Edmund
Bertschinger, Robert Crittenden, Catherine Heymans, Dragan Huterer,
Justin Khoury,  Douglas Scott and Ludovic Van Waerbeke for useful
comments and discussions, and Antony Lewis for help with CAMB. The
work of LP and GBZ was supported by a Discovery grant from NSERC and
by funds from SFU. The work of AS is supported, in part, by the
grant NSF- PHY-0653563 at Syracuse University and NSF AST-0708501 at
MIT. The work of JZ is supported by funds from NSERC, SFU, the J.
William Fulbright Foundation, and the University of California. The
authors gratefully acknowledge the use of facilities at the
Interdisciplinary Research in the Mathematical And Computational
Sciences (IRMACS) center at SFU.

\appendix
\section{Implementation of modified growth in CAMB and superhorizon consistency check}
\label{camb_implementation}

Following the notation in~\cite{MB}, we can map Eqs.~(\ref{parametrization-Poisson}) and
(\ref{parametrization-anisotropy}) into synchronous gauge by using
the following transformation,
\ba\label{gauge_trans} &&\Psi=\dot{\alpha}+\mathcal{H}\alpha~,\\
&&\Phi=\eta-\mathcal{H}\alpha~,\\
&&\alpha=(\dot{h}+6\dot{\eta})/2k^{2}~.\ea
Now, in the synchronous gauge, Eqs.~(\ref{parametrization-Poisson}) and
(\ref{parametrization-anisotropy}) become
\ba\label{eq:syn1}
&&k^{2}(\dot{\alpha}+\mathcal{H}\alpha)=-\frac{a^{2}}{2M_{p}^{2}}\mu\rho\Delta~,\\
\label{eq:syn2}
&&\frac{\eta-\mathcal{H}\alpha}{\dot{\alpha}+\mathcal{H}\alpha}=\gamma~.
\ea
Note that $\rho\Delta$ is gauge invariant. In the synchronous gauge, we
have \ba\label{comov_delta} &&\rho\Delta=\rho_{\rm m}\delta_{\rm
m}^{\rm s} ~.\ea and $\rho_{\rm m}=\rho_{\rm b}+\rho_{\rm
c},~\rho_{\rm m}\delta_{\rm m}^{\rm s}=\rho_{\rm b}\delta_{\rm
b}^{\rm s}+\rho_{\rm c}\delta_{\rm c}^{\rm s}$, where subscripts $\rm
m$, $\rm b$ and  $\rm c$ denote total matter, baryons, and cold dark
matter, respectively, and the superscript $\rm s$ denotes the
variables in synchronous gauge. Here and throughout, we assume that
baryons comove with CDM in the late universe where we modify gravity.

In addition, we have the energy-momentum conservation equations for
cold dark matter and baryons~\cite{MB}, \ba\label{eq:e-m_consv1}
&&\dot{\delta_{\rm c}^{\rm s}}=-\frac{1}{2}\dot{h}=3\dot{\eta}-k^{2}\alpha~,\\
\label{eq:e-m_consv2}&&\dot{\delta_{\rm b}^{\rm s}}=\dot{\delta_{\rm
c}^{\rm s}}~,\ea

To solve the set of coupled differential equations~(\ref{eq:syn1}),
(\ref{eq:syn2}), (\ref{eq:e-m_consv1}) and (\ref{eq:e-m_consv2}) ,
we start by eliminating $\dot{\eta}$ in Eq.~(\ref{eq:e-m_consv1}).
From Eq.~(\ref{eq:syn2}), we get, \ba\label{eta}
&&\eta=\gamma\dot{\alpha}+(1+\gamma)\mathcal{H}\alpha~. \ea Taking
the derivative w.r.t conformal time, \ba\label{etadot}
&&\dot{\eta}=\gamma\ddot{\alpha}+[(1+\gamma)\mathcal{H}+
\dot{\gamma}]\dot{\alpha}+[(1+\gamma)\dot{\mathcal{H}}+\mathcal{H}\dot{\gamma}]\alpha~.
\ea To obtain $\dot{\alpha}$ and $\ddot{\alpha}$ we note that
$\dot{\alpha}$ is given by Eq.~(\ref{eq:syn1}), \ba\label{alphadot}
&&\dot{\alpha}=-\mathcal{H}\alpha-\frac{a^{2}}{2M_{p}^{2}k^{2}}
\mu\rho\Delta~.\ea So, \ba\label{alphaddot}
&&\ddot{\alpha}=-\mathcal{H}\dot{\alpha}-\dot{\mathcal{H}}\alpha-\frac{\rho_{\rm
m}a^{2}}{2M_{p}^{2}k^{2}} [\mu\dot{\delta_{\rm m}^{\rm
s}}+\delta_{\rm m}^{\rm s} (\dot{\mu}-\mathcal{H}\mu)]~. \ea
Substituting Eq.~(\ref{alphadot}) and (\ref{alphaddot}) into
Eq.~(\ref{etadot}) and Eq.~(\ref{etadot}) into
Eq.~(\ref{eq:e-m_consv1}) and doing some algebra, we can finally get
the differential equation for $\delta_{c}^{s}$, which can be solved
by using the $\dot{\alpha}$ equation (\ref{alphadot}). In summary,
after redshift $30$, when gravity gets modified according to
Eqs.~(\ref{parametrization-Poisson}) and
(\ref{parametrization-anisotropy}), the equations we evolve
are\footnote{One should bear in mind that although cold dark matter
and baryons evolve according to the same differential equation in
the late universe, they have different initial conditions at
redshift $30$ when gravity gets modified. That's why we evolve them
separately in MGCAMB.}, \ba\label{eq:deltadot} &&\dot{\delta_{\rm
c}^{\rm s}}=\dot{\delta_{\rm b}^{\rm s}}=\dot{\delta_{\rm m}^{\rm
s}}=\f{-3a^{2}\rho_{\rm m}\delta_{\rm m}^{\rm s}
[\mu\dot{\gamma}+{\gamma}\dot{\mu}-\mathcal{H}\mu(\gamma-1)]+6M_{p}^{2}{\alpha}k^{2}(\dot{\mathcal{H}}-\mathcal{H}^{2})-
2M_{p}^{2}{k^4}\alpha}{(2M_{p}^{2}k^{2}+3\mu\gamma\rho_{\rm m}a^{2})},\\
&&\label{eq:alphadot}k^{2}\dot{\alpha}=-\mathcal{H}(k^{2}\alpha)-
\frac{\mu a^{2}\rho\Delta}{2M_{p}^{2}}~.\ea

In our modified version of CAMB, we take the values of $\delta_{\rm
c}^{\rm s},\delta_{\rm b}^{\rm s}~\rm{and}~\alpha$ from GR at redshift
$30$ as the initial conditions to evolve Eqs.~(\ref{eq:deltadot}) and (\ref{eq:alphadot}) at $z<30$.

Given $\alpha, \dot{\alpha}$ and $\delta_{\rm m}^{\rm s}$, we have all the
ingredients to calculate the observables, i.e the $C_\ell$'s. For example,
\ba\label{t1} &&\dot{h}=-2\dot{\delta_{\rm m}^{\rm s}}~,\\
\label{t2}&&\eta=\gamma\dot{\alpha}+\alpha\mathcal{H}(\gamma+1)~,\\
\label{t3}&&\dot{\eta}=(\dot{\delta_{\rm m}^{\rm s}}+k^{2}\alpha)/3~,\\
\label{t4}&&\Psi=-\frac{a^{2}\mu}{2M_{p}^{2}k^{2}}\rho\Delta~,\\
\label{t5}&&\Phi+\Psi=(1+\gamma)(\dot{\alpha}+\mathcal{H}\alpha)~,\\
\label{t6}&&\dot{\Psi}=-\frac{a^{2}\rho_{\rm m}\mu}{2M_{p}^{2}k^{2}}[\dot{\delta_{\rm m}^{\rm s}}+\delta_{\rm m}^{\rm s}(\frac{\dot{\mu}}{\mu}-\mathcal{H})]~,\\
\label{t7}&&\dot{\Phi}+\dot{\Psi}=\dot{\Psi}(1+\gamma)+\dot{\gamma}\Psi~.\ea

As described in~\cite{Bertschinger:2006aw}, if the theory is metric
based, obeys causality in the infrared limit, and the energy-momentum
conservation holds, then the evolution of super-horizon
perturbations is uniquely determined if the relation between the two
gravitational potentials is specified. Furthermore, in the long
wavelength limit, $k\rightarrow 0$, the ratio between the potentials
will be scale-independent, i.e. $\gamma\rightarrow\gamma(a)$. It is
important  to test whether the equations
(\ref{eq:deltadot}) and (\ref{eq:alphadot})~
 satisfy this consistency constraint on super-horizon scales. The
evolution of the metric potentials on super-horizon scales is
determined by the conservation equation for curvature perturbations~\cite{Bertschinger:2006aw,Hu:2007pj}. The relevant equations for the
super-horizon dynamics are \ba\label{consistency}
&&\Phi''+\Psi''-\f{H''}{H'}\Phi'+\l(\f{H'}{H}-\f{H''}{H'}\r)\Psi=0\,,\nonumber\\
&&\f{\Phi}{\Psi}=\gamma\,,
\ea
where a prime, just in this section, denotes the derivative with respect to $\ln{a}$, and $H\equiv\f{1}{a}\f{da}{dt}$.

We can combine the equations~(\ref{consistency}) into a second order
differential equation for the potential $\Psi$
\be\label{consistency_Psi}
\Psi''+\l(2\f{\gamma'}{\gamma}-\f{H''}{H'}+\f{1}{\gamma}\r)\Psi'+\l[\f{\gamma''}{\gamma}-\f{H''}{H'}\f{\gamma'}{\gamma}+\l(\f{H'}{H}-\f{H''}{H'}\r)\f{1}{\gamma}\r]\Psi=0\,.
\ee We shall check whether equations
(\ref{eq:deltadot}) and (\ref{eq:alphadot}) 
are consistent with Eq.~(\ref{consistency_Psi}). In order to do so, we
rewrite the equations
(\ref{eq:deltadot}) and (\ref{eq:alphadot}) 
 with independent variable
$\ln{(a)}$  and we combine them into a single first order
differential equation for $\Psi$ \be\label{eq_Psi} \Psi'=\f{(\mu\rho
a^2)'}{\mu\rho a^2}\Psi-\f{3\mu\rho a^2}{3\mu\rho a^2\gamma
+2M_P^2k^2}\l[\l(1+\f{(\mu\rho a^2\gamma)'}{\mu\rho
a^2}\r)\Psi+aH'\alpha-k^2\alpha\r] \ee In the super-horizon limit we
can neglect the term $2M_P^2k^2\ll3\mu\rho a^2\gamma$ for any $\mu$
and $\gamma$ which tend to a scale-independent finite function on
large scales and obtain \be\label{eq_Psi2}
\Psi'=-\f{1+\gamma'}{\gamma}\Psi-aH'\f{\alpha}{\gamma}\,. \ee Now,
we can take the derivative of~(\ref{eq_Psi2}) w.r.t. $\ln{a}$ to
obtain a second order differential equation for $\Psi$
\be\label{eq_Psi3}
\Psi''+\f{1+\gamma'}{\gamma}\Psi'+\l[\f{\gamma''}{\gamma}-\f{\gamma'}{\gamma}\f{1+\gamma'}{\gamma}\r]\Psi+\l[aH''-aH'\l(\f{\gamma'}{\gamma}-1\r)\r]\f{\alpha}{\gamma}+aH'\f{\alpha'}{\gamma}=0\,.
\ee
Finally, using the following equations
\ba\label{relation_alpha}
&&aH'\f{\alpha'}{\gamma}=\f{H'}{H}\f{\Psi}{\gamma}-aH'\f{\alpha}{\gamma}\nonumber\\
&&aH'\alpha\simeq -(1+\gamma')\Psi-\gamma\Psi'
\ea
it is easy to show that (\ref{eq_Psi3}) is equivalent to equation (\ref{consistency_Psi}). Therefore our equations give a consistent evolution on super-horizon scales.

\section{Scalar-tensor theories: mapping from the Einstein to the Jordan frame}\label{E_J_mapping}
The action for scalar-tensor theories in the Einstein frame reads
\be\label{Einstein_action}
S_E=\int d^4x\sqrt{-\tilde{g}}\l[\f{M_P^2}{2}\tilde{R}-\f{1}{2}\tilde{g^{\mu\nu}}(\tilde{\nabla}_{\mu}\phi)\tilde{\nabla}_{\nu}\phi-V(\phi)\r]+S_i\l(\chi_i,e^{-\kappa\alpha_i(\phi)}\tilde{g}_{\mu\nu}\r)\,,
\ee
where $\chi_i$ are the matter fields. $\phi$ is a scalar field, and $\alpha_i(\phi)$ represents the coupling of the scalar field to the $i$-th matter component. In what follows, we adopt the usual convention of indicating the Einstein frame quantities with a tilde.
The inverse of the conformal transformation~(\ref{mapping}) maps the theory to the Jordan frame, where matter falls along the geodesics of the metric while the Einstein-Hilbert action is modified. Using this conformal map, we can easily determine the Jordan frame quantities in terms of the corresponding Einstein ones. At the background level, for the scale factor and the component of the energy-momentum tensor, we have
\ba\label{background_map}
&&{\tilde a}^2= e^{\kappa\alpha_i(\phi)} \,a^2\hspace{1.2cm}{\tilde \rho}= e^{-2 \kappa\alpha_i(\phi)} \rho\,,\nonumber\\
&&{\tilde U}= e^{\kappa\alpha_i(\phi)/2} \,U\hspace{1cm}{\tilde P}= e^{-2 \kappa\alpha_i(\phi)} P\,,
\ea
where $U$ is the fluid $4$-velocity and $\rho$ and $P$ are, respectively, the fluid energy-density and pressure.
At the linear level, in the Newtonian gauge, we have
\ba\label{linear_mapping}
&&\tilde{\Psi}=\Psi+\f{\kappa\alpha_i'\delta\phi}{2},\hspace{1cm}\tilde{\Phi}=\Phi-\f{\kappa\alpha_i'\delta\phi}{2}\\
&&\tilde{\delta}=\delta-2\kappa\alpha_i'\delta\phi,\hspace{1cm}\tilde{\delta P}=\delta P - 2\kappa\alpha_i'\delta\phi\\
&&\tilde{v}=v,\hspace{2.5cm}\tilde{\sigma}=\sigma\,,
\ea
where the prime denotes a derivative w.r.t. the field $\phi$.

\section{Angular power spectra}
\label{append:cl}

Comparing predictions of cosmological models to observations involves projecting three-dimensional (3D) fields, such as the matter distribution or the gravitational potentials, onto two-dimensional (2D) fields measured by an observer looking at the sky.
Consider, for example, a 3D field ${\cal X}({\mathbf x},z)$, where $\mathbf{x}$ is the comoving coordinate and $z$ is the redshift (which we also use as our time variable). It would be observed via its weighted projection -- a 2D field $X(\mathbf{\hat{n}})$, which can be written as an integral along the line of sight:
\begin{equation}
X(\mathbf{\hat{n}}) = \int_0^{\infty} dz \ W_X(z) \ {\cal X}(\mathbf{\hat{n}}r(z),z) \ ,
\label{eq:field}
\end{equation}
where $\mathbf{\hat{n}}$ is a direction on the sky, $r(z)$ is the comoving distance to a point at redshift $z$, and $W_X(z)$ is a weight function, specific to the measurement, which selects a range along the radial coordinate that contributes to the observable $X(\mathbf{\hat{n}})$.

Ultimately, we are interested in the two-point correlation functions $C^{XY}(\theta) \equiv C^{XY}(|\mathbf{\hat{n}}_1 - \mathbf{\hat{n}}_2|) \equiv \left<X(\mathbf{\hat{n}}_1)Y(\mathbf{\hat{n}}_2)\right> $ between two fields -- the auto-correlations (where $X=Y$) and cross-correlations (where $X\neq Y$). For convenience, we work with the Fourier transform of the sources
\begin{equation}
{\cal X}(r\mathbf{\hat{n}},z) = \int \frac{d^{3}k}{(2\pi)^3} {\cal X}(\mathbf{k},z)e^{i\mathbf{k}\cdot \mathbf{\hat{n}}r}.
\label{eq:Fourier}
\end{equation}
The correlation function then becomes
\begin{equation}
C^{XY}(\theta) = \int d z_1 W_X(z_1) \int d z_2 W_Y(z_2) \int \frac{d^{3}k}{(2\pi)^3} e^{i\mathbf{k}\cdot \mathbf{\hat{n}}_1 r(z_1)} \int \frac{d^{3}k'}{(2\pi)^3} e^{i\mathbf{k'}\cdot \mathbf{\hat{n}}_2 r(z_2)} \left< {\cal X}(\mathbf{k},z_1){\cal Y}(\mathbf{k'},z_2) \right>.
\label{eq:Fourier_Corr}
\end{equation}

The temporal evolution of each Fourier mode depends only on the magnitude of its k-vector. Thus, we can write the source as a product of two factors. One contains the directional dependence at some early epoch $z_*$, deep in the radiation era, when all modes of interest were well outside the horizon. The other factor represents the time dependence and hence controls the evolution of our source with redshift:
\be
{\cal X}(\mathbf{k},\eta) \equiv {\cal X}(\mathbf{k}, z_*) \tilde{\cal X}(k,z) \ ,
\ee
where $\tilde{\cal X}(k,z_*)=1$. Then,
\be
\left< {\cal X}(\mathbf{k},z_1){\cal Y}(\mathbf{k'},z_2) \right> = \tilde{\cal X}(k,z_1) \tilde{\cal Y}(k',z_2) \left< {\cal X}(\mathbf{k},z_*){\cal Y}(\mathbf{k'},z_*) \right> \ .
\ee
At the high redshift $z_*$, when the modes are well outside the horizon and the modifications of gravity are negligible, the sources ${\cal X}(\mathbf{k},z_*)$ and ${\cal Y}(\mathbf{k},z_*)$ can be expressed in terms of the comoving curvature perturbation ${\cal R}$~\cite{liddlelythbook}. Namely, one can always introduce coefficients $c_{{\cal X}{\cal R}}$ and $c_{{\cal Y}{\cal R}}$ such that ${\cal X}(\mathbf{k},z_*) = c_{{\cal X}{\cal R}} {\cal R}(\mathbf{k},z_*)$ and similarly for ${\cal Y}$. Then, we can write
\be
\left< {\cal X}(\mathbf{k},z_*){\cal Y}(\mathbf{k'},z_*) \right> = c_{{\cal X}{\cal R}}c_{{\cal Y}{\cal R}} \left< {\cal R}(\mathbf{k},z_*){\cal R}(\mathbf{k'},z_*) \right>
\ee
The homogeneity of space implies that
\be
\left< {\cal R}(\mathbf{k},z_*){\cal R}(\mathbf{k'},z_*) \right> = (2 \pi)^3 \delta^{(3)}(\mathbf{k} + \mathbf{k'}) P_{\cal R}(k) \ ,
\ee
where $P_{\cal R}(k)$ is the primordial curvature power spectrum. One can also introduce the dimensionless spectrum, $\Delta_{\cal R}^{2} \equiv k^3  P_{\cal R} / 2 \pi^2$, and write
\begin{equation}
\left< {\cal X}(\mathbf{k},z_*){\cal Y}(\mathbf{k'},z_*) \right> = c_{{\cal X}{\cal R}}c_{{\cal Y}{\cal R}}(2 \pi)^3 \delta^{(3)}(\mathbf{k} + \mathbf{k'}) {2 \pi^2 \Delta_{R}^{2} \over k^3}
\label{eq:k-space_correlator}
\end{equation}
The correlation function $C^{XY}(\theta)$ can then be written as
\begin{equation}
C^{XY}(\theta) = c_{{\cal X}{\cal R}}c_{{\cal Y}{\cal R}} \int_0^{z_*} d z_1 W_{\cal X}(z_1) \int_0^{z_*} d z_2 W_{\cal Y}(z_2) \int \frac{d^3 k}{4 \pi k^3} \Delta_{R}^{2} e^{i\mathbf{k}\cdot(\mathbf{\hat{n}}_1r_1 - \mathbf{\hat{n}}_2r_2)}\tilde{\cal X}(k,z_1) \tilde{\cal Y}(k,z_2)
\end{equation}
Integrating over the directional dependence of ${\mathbf  k}$, and defining $R = \sqrt{r_1^2 + r_2^2 - 2r_1r_2\cos\theta}$, we obtain
\begin{equation}
C^{XY}(\theta) = c_{X\Psi}c_{Y\Psi} \int_0^{z_*} d z_1 W_{\cal X}(z_1) \int_0^{z_*} d z_2 W_{\cal Y}(z_2) \int \frac{dk}{k} \Delta_{R}^{2} \frac{\sin{kR}}{kR} \tilde{\cal X}(k,z_1) \tilde{\cal Y}(k,z_2).
\label{eq:sinc_integ}
\end{equation}
We can further expand Eq.~(\ref{eq:sinc_integ}) in a Legendre series:
\begin{equation}
C^{XY}(\theta) = \sum_{\ell=0}^{\infty} \frac{2\ell+1}{4 \pi}C_\ell^{XY} P_\ell(\cos\theta) \ ,
\label{eq:series}
\end{equation}
where the first two terms in the series (the monopole and dipole contributions) are coordinate dependent and should vanish in the CMB frame of a homogeneous and isotropic universe. Expanding $\sin(kR)/kR$ in terms of spherical Bessel functions $j_\ell$, and comparing the result with Eq.~(\ref{eq:series}), we can write
\begin{equation}
C_\ell^{XY}= 4\pi \int \frac{dk}{k} \Delta_{\cal R}^{2}  I_{\ell}^X(k) I_{\ell}^Y(k),
\label{eq:C_gen}
\end{equation}
where
\begin{equation}
I_{\ell}^X(k) =  c_{{\cal X}{\cal R}}\int_0^{z_*} d z W_X(z)  j_\ell[kr(z)]\tilde{\cal X}(k, z).
\label{eq:I_gen_2}
\end{equation}
and similarly for $I_\ell^Y$. The expressions for $I_\ell$'s for galaxy distributions in redshift bins, the maps of lensing shear for different bins, and the CMB temperature are given in Sec~\ref{sec:cl}. One can then calculate various types of correlations between the different fields using Eq.~(\ref{eq:C_gen}).

\section{Optimal stepsize for finite difference derivatives}
\label{fisher_stepsize}

\begin{table*}[tb]
\begin{tabular}{cccccc}
\hline\hline
$\textbf{P}$&$\Omega_bh^{2}$ &$\Omega_ch^{2}$  &$h$    &$\tau$    &$n_s$\\
\hline
$\Delta{p}$ &0.005           &0.005            &0.03   &0.02      &0.005\\

\end{tabular}

\begin{tabular}{c|cccc}

\hline\hline Model   & I&{II}&{III} & {IV}\\
{\textbf{P}}&\multicolumn{4}{c}{$\Delta{p}$}\\

   \hline
$\lama$   &0.3    &0.4    &0.3    &0.4\\
$\beta_1$ &0.1    &0.1    &0.1    &0.1\\
$\lamb$   &0.3    &0.4    &0.3    &0.4\\
$\beta_2$ &0.05   &0.05   &0.1    &0.1\\
$s$       &0.4    &0.4    &0.2    &0.2\\
\hline \hline
\end{tabular}
\caption{The optimal stepsize for the cosmological
parameters we use in this work.}  \label{tab:stepsize}
\end{table*}

For the calculation of the Fisher information matrix, we need to take
numerical derivatives of our tomographic observables, say,
$C_{\ell}^{XY}$ w.r.t cosmological parameters $p$. For small
$\Delta{p}$, we can use the approximation \be\label{stepsize}
{\partial}C_{\ell}^{XY}/{\partial}p \sim
[C_{\ell}^{XY}(p+{\Delta}p)-C_{\ell}^{XY}(p-{\Delta}p)]/2{\Delta}p\ee
Note that ${\Delta}p$ cannot be too big, which makes the right-hand-side of Eq.~(\ref{stepsize}) deviate too much from the derivative, neither too
small, which might give rise to numerical instability. To get a
proper range for ${\Delta}p$ is a bit of guesswork. One often has to do numerical tests until some
range of ${\Delta}p$ is found so that ${\Delta}C_{l}^{XY}/{\Delta}p$
converges when ${\Delta}p$ varies in this range. In Table~\ref{tab:stepsize} we list the ${\Delta}p$ we found and used in this work.

\section{Reconstruction of $\mu$ and $\gamma$ by error
propagation}
 Given the variances of the five parameters listed in Tables
\ref{tab:DES} and \ref{tab:LSST}, namely, Var[$\lama$],
Var[$\beta_{1}$], Var[$\lamb$], Var[$\beta_{2}$], Var[$s$],  and the
corresponding covariance matrices, one can estimate the variance of
$\mu$ and $\gamma$ at given $k$ and $z$ by error propagation.
$\mu$ and $\gamma$ have the same functional form:
\ba\label{eq:f} f&=&\frac{1+{\beta}t}{1+t},\\
\label{eq:t} t&=&{\Lambda}(k\cdot{\rm Mpc})^{2}a^{s},\ea By
perturbing Eq.~(\ref{eq:f}) and (\ref{eq:t}), we have
\ba\label{eq:pert_f}
{\Delta}f&=&[(\beta-f){\Delta}t+t{\Delta}\Lambda]/(1+t),\\
\label{eq:pert_t}{\Delta}t&=&t[{\rm ln}10\cdot\Delta({\rm
log\Lambda})+{\rm ln}a\cdot\Delta{s}],\ea Note that here we make
$\Lambda$ dimensionless for the ease of taking the logarithm later.
So $\Lambda$ is basically equivalent to $\lama$ or $\lamb$.
Taking the square of Eqs.~(\ref{eq:pert_f}) and (\ref{eq:pert_t})
and plugging in the corresponding entries of the covariance matrix,
we finally get \ba\label{eq:var_f} {\rm
Var}[f]&=&\Big(\frac{t}{1+t}\Big)^{2}\Big\{[{\rm
ln10}\cdot(\beta-f)]^{2}{\rm Var}[{\rm log\Lambda }] +
[{\rm ln}a\cdot(\beta-f)]^{2}{\rm Var}[s]+{\rm Var}[\beta] \nonumber\\
&&+2(\beta-f)^2{\rm ln}10\cdot{\rm ln}a\cdot{\rm Cov}[{\rm
log\Lambda},s] \nonumber\\
&&+2(\beta-f){\rm ln}10\cdot{\rm Cov[{\rm log\Lambda},\beta]}\nonumber\\
&&+2(\beta-f){\rm ln}a\cdot{\rm Cov}[s,\beta] \Big\}. \ea where
Cov[$A,B$] denotes the covariance of two random variables $A$ and
$B$, and the square root of Var$[f]$ gives the errors on $\mu$ and
$\gamma$.


\end{document}